\documentclass[preprint,12pt]{elsarticle}
\usepackage{amssymb,amsfonts,amsthm}
\usepackage{geometry}
\usepackage{graphics,subfigure}
\usepackage{booktabs}
\usepackage{hyperref}
\usepackage{pifont}
\usepackage[fleqn]{amsmath}
\usepackage{color}
\usepackage{lineno}
\usepackage{bm}
\numberwithin{equation}{section}

\begin{document}
\title{Surface wrinkling of a film coated to a graded substrate}

\author[add1]{Rui-Cheng Liu}
\author[add1,add2,add3]{Yang Liu\corref{cor1}}
\ead{tracy\_liu@tju.edu.cn}
\author[add3]{Alain Goriely\corref{cor1}}
\ead{goriely@maths.ox.ac.uk}

\address[add1]{Department of Mechanics, School of Mechanical Engineering, Tianjin University, Tianjin 300354, China}
\address[add2]{Tianjin Key Laboratory of Modern Engineering Mechanics, Tianjin 300354, China}
\address[add3]{Mathematical Institute, University of Oxford, Oxford, OX2 6GG, UK}

\cortext[cor1]{Corresponding author.}
  
\begin{abstract}
We study the surface wrinkling of a stiff thin elastic film bonded to a compliant graded elastic substrate subject to compressive stress generated either by compression or growth of the bilayer. Our aim is to clarify the influence of the modulus gradient on the onset and surface pattern in this bilayers. Within the framework of finite elasticity, an exact bifurcation condition is obtained using the Stroh formulation and the surface impedance matrix method. Further analytical progress is made by focusing on the case of short wavelength limit for which the Wentzel-Kramers-Brillouin method can be used to resolve the eigenvalue problem of ordinary differential equations with variable coefficients. An explicit bifurcation condition is  obtained from which asymptotic the critical buckling load and the critical wavelength are derived. In particular, we consider two distinct situations depending  on  the ratio $\beta$ of the shear modulus at the substrate surface to that at infinity. If $\beta$ is of $\mathcal{O}(1)$ or small,  the parameters related to modulus gradient all appear in the high order terms and play an insignificant role in the bifurcation. In that case, it is the modulus ratio between the film and substrate surface that governs the onset of surface wrinkling. If, however, $\beta\gg1$, the modulus gradient  affects the critical condition through leading-order terms. Through our analysis we unravel the influence of different material and geometric parameters, including the modulus gradient, on the bifurcation threshold and the associated wavelength which can be of importance in many biological and technological settings.
\end{abstract}


\maketitle  

\section{Introduction}
The wrinkling of thin layers on substrates is a universal morphological phenomenon of great importance  in engineering materials and biological tissues  \citep{Li2012SM,Gorielybook}. A typical set-up to study wrinkling is to consider a thin hard film coated to a soft substrate. If the film shrinks, grows, is prestrained, or the entire system is loaded, the film may develop wrinkles \cite{Alawiye2019PTRSA}.   This type of wrinkling morphologies appear in biological systems and is known to establish some of their physiological functions such as  folds in human brain \citep{Budday2014,hobugo18,Balbi2020PRE} or  in gastrointestinal tracts and airway walls \citep{Balbi2015JMPS,Eskandari2013,Eskandari2016}. From the viewpoint of mechanics, wrinkled or buckled shapes in skins, fruits or vegetables can be viewed as a result of a bifurcation process related to  growth or aging  \citep{Kucken2004EPL,Yin2010AB,Dai2014EPL}. In engineering, wrinkling in film/substrate bilayers can be used to design specific patterns that  can be used to alter the optical property \citep{van2013AM,Li2017,Lee2010AM}, to regulate surface adhesion or tension \citep{Chan2008AM,Yang2010AFM,Lee2021EML}, to optimize the wetting characteristic  \citep{Zhang2012JACS,Sabbah2016AS}, to assist in measuring the material properties of soft polymer networks \citep{Stafford2004NM,Wilder2006JACS,Cao2009,Chan2009}, or to help design novel flexible sensors \citep{Zhao2020Langmuir,Wang2021PAT,Lee2022Small}.  Moreover, certain surface morphologies brought by structure buckling are expected to occur and will play a positive role in some special cases, such as the creation of gecko's palm surfaces with adhesive force \citep{Autumn2002PNAS}. Therefore, it is of great significance to explore the buckling and post-buckling of film/substrate bilayers to understand and further to control multiple pattern formations.

%

The fundamental importance and  potential applications of film/substrate bilayers  have led to a flourish of work related to instability problem when, typically, both  film and substrate are  homogeneous materials. Early works were dedicated to determining the onset of surface instabilities induced by axial compression \citep{Dorris1980JAM,Shield1994JAM,Ogden1996,Bigoni1997IJSS,Steigmann1997PRSA,Cai2000IJSS}. Within the framework of nonlinear elasticity, \cite{Cai1999PRSA} conducted both the linear and weakly non-linear analyses for a compressed neo-Hookean bilayer. The main finding is that there exists a critical value of the modulus ratio between two layers, across which a transition between supercritical and subcritical bifurcations occurs. Later, the influence of compressibility \citep{Liu2014IJES,Cai2019IJNLM}, curvature \cite{jia2018curvature}, pre-stretch \citep{Song2008IJSS,Hutchinson2013PTRSA}, growth \citep{Li2011JB,Li2011JMPS,Budday2015,Moulton2011JMPS}, multi-layering \citep{Cheng2014IJSS,Wang2020IJSS,Zhou2022MMS}, viscosity \citep{Huang2002IJSS,Huang2005JMPS}, fibers \cite{stewart2016wrinkling}, and anisotropy \citep{Im2008JMPS,Nguyen2020BMM} were addressed theoretically, numerically and experimentally. In particular, \cite{Fu2015PRSA} studied the buckling and post-buckling of a film-substrate bilayer when they have similar material properties, and \cite{Hutchinson2013PTRSA} extended the analysis in \cite{Cai1999PRSA} by considering the pre-stretch effect. Recently, \cite{Alawiye2019PTRSA,Alawiye2020JMPS} revisited this classical problem as well and further explored the growth induced surface instability for which higher-order asymptotic solutions for the critical buckling loads and the amplitude equation were obtained.

 
In many  systems, such as human airway wall \citep{Eskandari2016,Liu2020IJNLM} and hydrogels with graded cross-linking density and surface modified polydimethylsiloxane (PDMS) \citep{Qi2018ACS,Khare2009JACS,Wang2016APL}, the elastic modulus or Poisson's ratio of the substrate may be position dependent due to the spatial variation in the microstructure, leading to a graded substrate. Previous studies have looked at  the initiation of surface buckling for an elastic graded half-space or block\citep{Lee2008JMPS,Wu2013IJSS,Wu2014JAM,Yang2017PRSA,Diab2014PRSA,Chen2018}. In particular \cite{Yang2017PRSA} adopted an exponential modulus distribution and obtained an explicit bifurcation condition for a compressed block. Yet such choice for modulus functions lead to zero modulus at infinity in  a half-space. 
For film/substrate bilayers, \cite{Cao2012IJSS} investigated surface instability of a stiff film on an elastic graded half-space. They derived an analytical solution of the critical compressive strain and the critical wavelength for power-law grading modulus and exponential grading modulus, respectively and the post-buckling solution was studied numerically. \cite{Jia2014IJAM} established a semi-analytical finite element model to obtain the bifurcation threshold for an elastic graded cylinder covered by a stiff film under axial compression. \cite{Chen2017PRSA} further developed an analytical model for the buckling of a stiff film on a graded half-space and performed a finite element analysis to validate the analytical model as well as to investigate the post-buckling behavior.

 When the material has a graded structure, the governing equations include space-dependent variable coefficients which is particularly difficult to study analytically. For this reason, most  studies have resorted to  numerical methods to solve the bifurcation problem.
 The present work is interested in obtain analytical methods for the wrinkling problem in graded material in order to obtain a comprehensive picture on the role of  geometrical and material parameters. 
Here, we use the fact that for thin stiff film resting on a soft graded substrate, the critical wavelength is small compared to the length of the film, indicating a large critical wavenumber \citep{Cao2012IJSS,Chen2017PRSA}. Hence, we can use the WKB Method (WKB stands for Wentzel-Kramers-Brillouin) \citep{Hinch1991} to constructing asymptotic solutions for ordinary differential equations with small parameters. It has been used by \cite{Fu1998IJNLM,BenAmar2005JMPS} to derive asymptotic results of buckling of spherical shells of arbitrary thickness. After that this asymptotic  method was successfully applied to deal with the bifurcation analysis arising from the problems of cylindrical tubes \citep{Sanjarani2002SIAM,Sanjarani2010IJAM,Liu2018IJNLM,Jin2018IJES}, spherical shells \citep{Haughton2003ZAMP,Jia2018PRE,DePascalis2023EML}, and bending of blocks \citep{Coman2008QJMAM,Sanjarani2013IJES}.





\begin{figure}[ht!]
    \centering\includegraphics[scale=0.23]{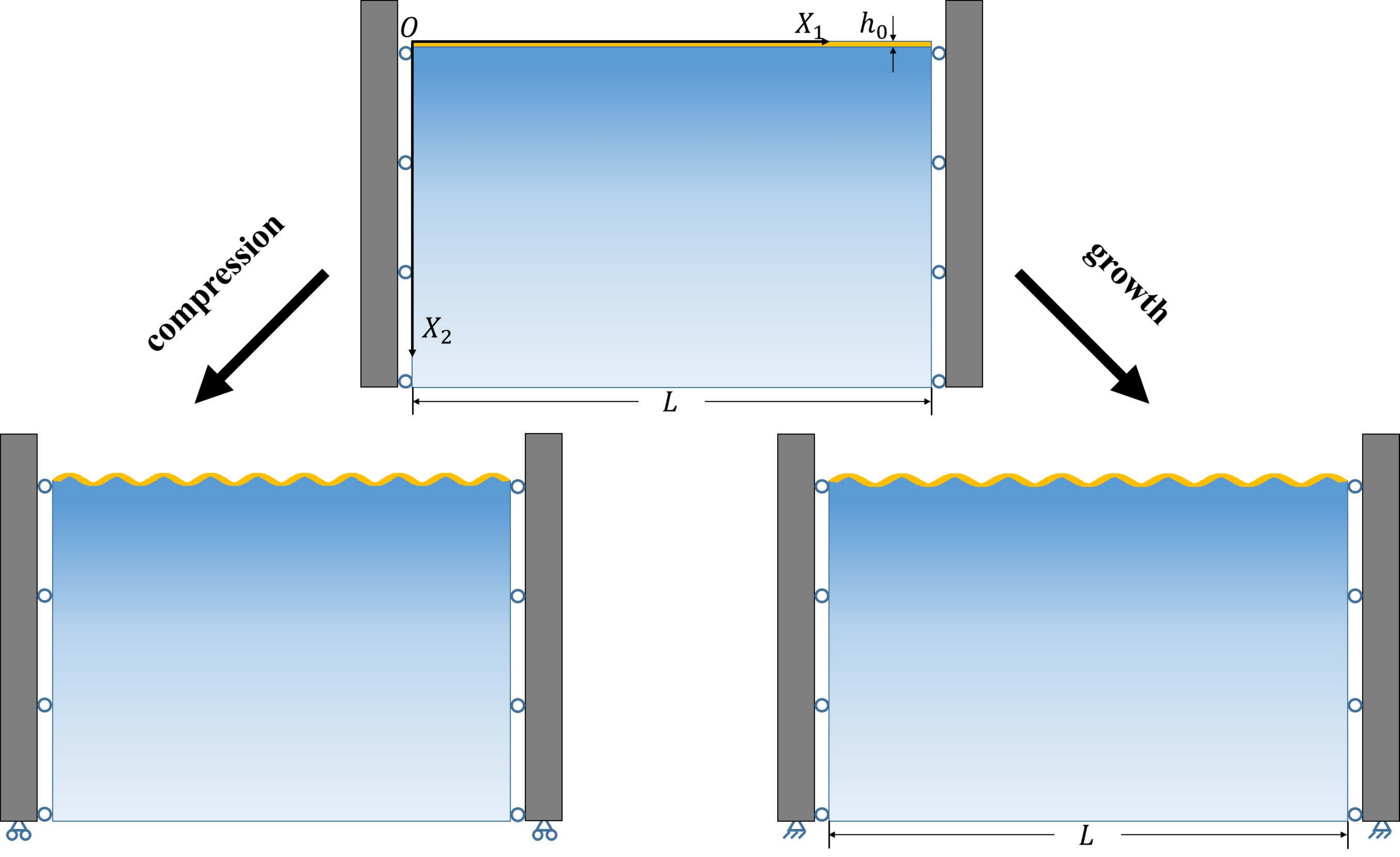}
    \caption{Schematic illustration of a film coated to a graded half-space under uniaxial compression or growth. The initial length of the bilayer is $L$ and the thickness of the film is $h_0$. In both loading cases, the sliding end conditions are used such that there is no shear stress components in the two ends. In addition, the length $L$ is fixed in the growth scenario so the stress is induced by restricted growth.}
    \label{geometry}
\end{figure}

In this paper, we work within the framework of finite elasticity and study how the modulus gradient affects the critical buckling condition of a bilayer system with a graded substrate. We obtain explicit form for the critical buckling stretch and the corresponding critical wavenumber for the case of a compressed block and the case of a growing bilayer as shown in Figure~\ref{geometry}. 
In Section \ref{Theoretical-model}, we characterize the primary deformation prior to surface instability and present the incremental theory under in-plane compression and restricted growth. A linear bifurcation analysis is carried out using the Stroh formulation and the surface impedance matrix method in Section \ref{Bifurcation-analysis}. We use WKB method to obtain an explicit bifurcation condition in Section \ref{WKB}. The asymptotic solutions are established in Section \ref{Asymptotic-analysis-decay} for $\beta$  small or of $\mathcal{O}(1)$ where $\beta$ is the ratio of the shear modulus of the surface of the substrate to that at infinity. The reverse case where $\beta$ is large is explored in Section \ref{Asymptotic-analysis-grow}. 


\section{Theoretical model}\label{Theoretical-model}

We consider a film/substrate bilayer  composed of incompressible hyperelastic materials under compression or growth. The  shear modulus of the substrate is assumed to decrease or increase uniformly in the thickness direction (away from the film). We assume that the bilayer is in a state of plane-strain deformation and therefore  restrict our attention to the planar problem as shown in Figure~\ref{geometry}. The origin $O$ is located at the  left end of the film surface, and the $X_1$- and $X_2$-axes correspond to the length and depth directions, respectively. We denote the initial length of the bilayer by $L$ and the initial thickness of the film by $h_{0}$. The substrate is assumed to be a half-space. Therefore, in the reference configuration the body $\mathcal{B}_{0}$ occupies a region $[0,L]\times \left([0,h_0] \cup [h_0,\infty)\right)$, and  any material point can be represented in the  orthonormal basis $\{\mathbf{e}_1, \mathbf{e}_2 \}$ by its initial position $\mathbf{X}=X_1 \mathbf{e}_1+X_2\mathbf{e}_2$. 

We assume that the film surface ($X_2=0$) is traction-free and that there is no displacement as $X_2\to\infty$. On the vertical sides, the bilayer is allowed to slide along but not penetrate the sides at $X=0$ and $L$. 
We assume perfect bonding of the interface ($X_2=h_{0}$) so that both  traction and displacement are continuous across this interface. 

When loads are applied through uniaxial either compression or growth the bilayer undergoes a finite deformation and the resulting configuration, denoted by $\mathcal{B}_{r}$, occupies the region $[0,l]\times \left( [0,h] \cup [h,\infty)\right)$ for the compression case or within the region $[0,L]\times \left([0,h] \cup [h,\infty)\right)$ for the growth case. The  position of a point originally at $\mathbf{X}$  is now  $\mathbf{x}({\mathbf X})=x_1\mathbf{e}_1+x_2\mathbf{e}_2$. 
The deformation gradient associated with the deformation $\mathcal{B}_{0} \rightarrow \mathcal{B}_{r}$ is given by
\begin{align}
    \mathbf{F}=\frac{\partial {\mathbf x}}{\partial {\mathbf X}}.
    \label{deformation-gradient}
\end{align}
As the applied external load reaches a critical value, surface wrinkling may emerge as a  bifurcation and this  state is referred to as $\mathcal{B}_{t}$. Since there are two layers in the model, we use an upper hat for any quantity belonging to the film and no hat otherwise. For example, $\hat{W}$ represents the strain-energy function for the film while the strain-energy function for substrate is written as $W$. Since the derivations of the governing equations are similar for both layers, we only show explicitly the computation for the substrate. The equations for film are obtained by changing variables.

We consider two loading scenarios, namely, uniaxial compression and restricted growth. In both cases, we will first characterize analytically the base state $\mathcal{B}_{r}$ and then derive the linearized incremental governing equations from $\mathcal{B}_{r} \rightarrow \mathcal{B}_{t}$.

\subsection{Uniaxial compression}\label{Uniaxial-compression}
In the compression case, the deformation gradient \eqref{deformation-gradient} for the base state $\mathcal{B}_{r}$ reduces to
\begin{align}
    \mathbf{F}=
    \lambda_1\mathbf{e}_1 \otimes \mathbf{e}_1+\lambda_2\mathbf{e}_2 \otimes \mathbf{e}_2,
    \label{deformation-gradient-compression}
\end{align}
where $\lambda_1$ and $\lambda_2$ are the principal stretches. The incompressibilily condition $\det \mathbf F=1$ implies $\lambda\equiv\lambda_1=\lambda_2^{-1}$ and the length and height of this deformed body are:
\begin{align}
l=\lambda L,\quad h=h_0/\lambda.
\end{align} 
Denoting the strain-energy function of the substrate by $W=W(\mathbf{F})$ or, with a slight abuse of notation, in terms of the principal stretches by $W(\lambda_1,\lambda_2)$, the Cauchy stress tensor $\bm \sigma$ in $\mathcal{B}_{r}$ is given by
\begin{align}
    \bm \sigma=\mathbf{F} \frac{\partial W}{\partial \mathbf{F}}-p \mathbf{I},
    \label{Cauchy-stress}
\end{align}
where $p$ is the Lagrange multiplier enforcing the incompressibility condition and $\mathbf{I}$ is the second-order identity tensor. Neglecting  body forces, the equilibrium equation for a static problem is:
\begin{align}
    \operatorname{div}\bm{\sigma}=\mathbf{0},
    \label{equation-compression}
\end{align}
where  `$\operatorname{div}$' is the divergence operator  in the current configuration. From the traction-free boundary condition at $y=0$ and the  continuity condition on the interface $y=h$, we obtain
\begin{align}
   \hat{p}=\lambda^{-1} \hat{W}_{2}, \quad
    p=\lambda^{-1} W_2,
    \label{pressure}
\end{align}
where $W_2=\partial W / \partial \lambda_2$.

\subsection{Restricted growth}\label{Restricted-growth}
In this scenario, the length between the two rigid plates in Figure \ref{geometry} is fixed to be $L$ and the deformation is induced by growth. We refer to the theory of volumetric growth \citep{Gorielybook} that uses a multiplicative decomposition of \eqref{deformation-gradient}:
\begin{align}
    \mathbf{F}=\mathbf{A} \mathbf{G},
    \label{deformation-gradient-growth}
\end{align}
where $\mathbf{A}$ is an elastic deformation tensor and $\mathbf{G}$ is the growth tensor. We assume that both $\mathbf{A}$ and $\mathbf{G}$ are diagonal tensors such that 
\begin{align}
    \mathbf{A}=\alpha_1 \mathbf{e}_1 \otimes \mathbf{e}_1+ \alpha_2 \mathbf{e}_2 \otimes \mathbf{e}_2,\quad \mathbf{G}=g_1 \mathbf{e}_1 \otimes \mathbf{e}_1 +g_2 \mathbf{e}_2 \otimes \mathbf{e}_2,
    \label{eq2.7}
\end{align}
where $\alpha_i$ ($i=1, 2$) represent the elastic principal stretches, and $g_i$ are growth factors. There is no volume change in the $i$-direction when $g_i=1$, and $g_i>1$, $g_i<1$ represent growth and shrinkage, respectively. In this work, we only consider volume increase, meaning that  the growth factors are always greater than unity. Since the deformation in the $x$-direction is limited so that $\lambda_1=1$, we  have $\alpha_1=g^{-1}_1$. Further, according to the elastic incompressibility condition $\det\mathbf{A}=1$, we have $\alpha_1\alpha_2=1$ and hence $\alpha_2=g_1$ and $\lambda_2=g_1g_2$. Here we denote the strain-energy function by $W=W(\mathbf{A})$ (or $W=W(\alpha_1,\alpha_2)$), and an explicit expression of the Cauchy stress \eqref{Cauchy-stress} is given by \citep{Gorielybook}:
\begin{align}
    \bm \sigma=\mathbf{A} \frac{\partial W}{\partial \mathbf{A}}-p \mathbf{I}.
    \label{stress-growth}
\end{align}

According to the equilibrium equation \eqref{equation-compression}, the traction-free boundary condition on $y=0$, and the traction continuity condition on the interface $y=h$, we find
\begin{align}
    \hat{p}=\hat{g}_1 \hat{W}_2, \quad
     p=g_1 W_2,
    \label{pressure-growth}
\end{align}
where $W_2=\partial W/\partial\alpha_2$.

\subsection{Incremental theory}\label{Incremental-theory}
Next, we are interested in the possible existence of some states $\mathcal{B}_{t}$ close to $\mathcal{B}_{r}$. We obtain them by superimposing an infinitesimal displacement field $\mathbf{u}(\mathbf{x})=u_1(x_1,x_2)\mathbf{e}_1+u_2(x_1,x_2)\mathbf{e}_2$ on $\mathcal{B}_{r}$. In this way, the position vector of a material particle relative to the common orthonormal basis is denoted by $\tilde{\mathbf{x}}$, which satisfies the following relation
\begin{align}
    \tilde{\mathbf{x}}(\mathbf{X})=\mathbf{x}(\mathbf{X})+\mathbf{u}(\mathbf{x}).
    \label{relation}
\end{align}
The deformation gradient arising from $\mathcal{B}_{0} \rightarrow \mathcal{B}_{t}$ is $\tilde{\mathbf{F}}=\partial \tilde{\mathbf{x}} / \partial \mathbf{X}=(\mathbf{I}+\bm \eta)\mathbf{F}$, with
\begin{align}
    \bm \eta=\frac{\partial u_i}{\partial x_j} \mathbf{e}_i \otimes \mathbf{e}_j\equiv u_{i,j} \mathbf{e}_i \otimes \mathbf{e}_j,
    \label{displacement-gradient}
\end{align}
where repeated indices imply a summation from 1 to 2.
The linearized incompressibility constraint then reads
\begin{align}
    \operatorname{tr}\bm \eta=u_{i,i}=0,
    \label{linearized-incompressibility}
\end{align}
where `$\operatorname{tr}$' is the trace operator.

The first Piola-Kirchhoff stress tensors in $\mathcal{B}_r$ and $\mathcal{B}_t$ are denoted by $\bm\pi$ and $\tilde{\bm {\pi}}$, respectively. It is convenient to introduce the incremental stress tensor $\bm \chi$ by
\begin{align}
    \bm \chi= J^{-1} (\tilde{\bm {\pi}}-\bm {\pi}) \mathbf{F}^{\textrm{T}},
    \label{incremental-stress}
\end{align}
where $J=\det \mathbf{F}$ and `$\textrm{T}$' is the transpose. The incremental equations for the substrate in terms of $\bm \chi$ are
\begin{align}
    \textrm{div}~\bm \chi^{\textrm{T}}=\mathbf{0},
    \label{Incremental-equation}
\end{align}
and, similarly, for the film:
\begin{align}
    \textrm{div}~\hat{\bm \chi}^{\textrm{T}}=\mathbf{0}.
    \label{Incremental-equation2}
\end{align}

The  incremental boundary condition and continuity condition are:\\
(i) the traction-free boundary condition,
\begin{align}
    \hat {\bm \chi} \mathbf{e}_2=\mathbf{0}, \quad \textrm{on}~x_2=0;
    \label{Incremental-bc}
\end{align}
(ii) the continuity condition,
\begin{align}
    ({\bm \chi}-\hat {\bm \chi}) \mathbf{e}_2=\mathbf{0}, \quad \mathbf{u} -\hat{\mathbf{u}}=\mathbf{0}, \quad \textrm{on}~x_2=h;
    \label{Incremental-cc}
\end{align}
(iii) the decay condition,
\begin{align}
    \mathbf{u}=\mathbf{0}, \quad \textrm{as}~x_2\rightarrow \infty;
    \label{Incremental-dc}
\end{align}
(iv) the  sliding-end condition,
\begin{align}
    {\bm \chi}\mathbf{e}_1=0, \quad u_{1,2}=0,\quad \textrm{on}~x_1=0~\textrm{and}~l.
    \label{sliding-end}
\end{align}

For an incremental deformation, the magnitude of $\bm\eta$ is small and we can therefore expand $\bm\chi$ in $\bm \eta$ and retain the linear terms to obtain
\begin{align}
    \chi_{ij}=\mathcal{A}_{jilk} \eta_{kl}+p \eta_{ji}-\dot{p} \delta_{ji}+\mathcal{O}(|\eta_{ij}|^2),
    \label{expansion}
\end{align}
where the pressure $p$ has been defined in \eqref{Cauchy-stress}, $\dot{p}$ is the incremental counterpart, and the fourth-order tensor $\bm {\mathcal{A}}$ defines instantaneous moduli whose components are given by
\begin{align}
    \mathcal{A}_{jilk}=F_{jA} F_{lB} \frac{\partial^{2}W}{\partial \tilde{F}_{iA} \partial \tilde{F}_{kB}} \bigg|_{\tilde{\mathbf{F}}=\mathbf{F}}, \quad\textrm{in}~\textrm{the compression}~\textrm{case},
    \label{moduli-compression}
\end{align}
\begin{align}
    \mathcal{A}_{jilk}=A_{jA} A_{lB} \frac{\partial^{2}W}{\partial \tilde{A}_{iA} \partial \tilde{A}_{kB}} \bigg|_{\tilde{\mathbf{A}}=\mathbf{A}},\quad\textrm{in}~\textrm{the growth}~\textrm{case}.
    \label{moduli-growth}
\end{align}

The set of equations \eqref{Incremental-equation}-\eqref{sliding-end} forms a linear system for which we seek solutions. More precisely, we are interested in finding  values of the parameter (either compression or growth), such that there exists a non-trivial solution to this system. These values are the critical values for the bifurcation. 


\section{Bifurcation analysis}\label{Bifurcation-analysis}
We will apply the Stroh method \citep{Stroh1962JMP,Fu2002PRSA,Fu2005JE,Su2016IJSS,Su2019IJSS,Su2020IJES,Liu2022MMS} and the surface impedance matrix method \citep{Biryukov1985,Biryukov1995,Shuvalov2003PRSA,Shuvalov2003QJMAM} to obtain  the critical buckling load and its associated wavenumber rather than the determinant method or the compound matrix method \citep{Jin2018IJES,Liu2018IJNLM}. 


\subsection{Stroh formulation and the surface impedance matrix method}\label{Stroh}

To describe the initial wrinkling pattern, we look for a solution of the form
\begin{align}
    &u_1(x_1,x_2)=U(x_2)\sin(n x_1),\  u_2(x_2,x_2)=V(x_2)\cos(n x_1),\  \dot{p}(x_1,x_2)=P(x_2)\cos(n x_1),
    \label{Solution-form}
\end{align}
where $U(x_2)$, $V(x_2)$ and $P(x_2)$ are unknown functions to be determined and $n$ is the wavenumber. Note that in the current problem the same $n$ is used for both the substrate and the film. It then follows from the sliding end condition \eqref{sliding-end} that
\begin{align}
    n=\dfrac{k \pi}{l},
    \label{Wavenumber}
\end{align}
where $k$ is the dimensionless wavenumber and $l=\lambda L$. Remember that in the growth case the principal stretch $\lambda$ is always unity.

The components of the incremental stress tensor $\bm\chi$ can be expressed as 
\begin{align}
    \chi_{12}(x_1,x_2)=T_{12}(x_2)\sin(n x_1),\quad \chi_{22}(x_1,x_2)=T_{22}(x_2)\cos(n x_1),
    \label{Separation}
\end{align}
where $T_{12}(x_2)$ and $T_{22}(x_2)$ are unknown functions. 

To simplify book-keeping, we scale all coordinates and variables of length dimension by $L$. In addition, all variables of the stress unit are normalized by $\mu_h$. Here $L$ is the length of the bilayer and $\mu_{h}$ is the shear modulus on the substrate surface $(X_2=h_0)$ yielding
\begin{equation}
\begin{aligned}
&x =\dfrac{x_1}{L},\quad y =\dfrac{x_2}{L},\quad X =\dfrac{X_1}{L},\quad Y =\dfrac{X_2}{L},\quad U^* =\dfrac{U}{L},\quad V^* =\dfrac{V}{L},\quad h _{0}^*=\dfrac{h_{0}}{L}, \quad h^* =\dfrac{h}{L},\\
&p^* =\dfrac{p}{\mu_{h}},\quad P^* =\dfrac{P}{\mu_{h}},\quad \mathcal{A} _{jilk}^*=\dfrac{\mathcal{A}_{jilk}}{\mu_{h}},\quad T_{12}^* =\dfrac{T_{12}}{\mu_{h}},\quad T_{22}^* =\dfrac{T_{22}}{\mu_{h}},\quad \hat{\mathcal{A}}^* _{jilk}=\dfrac{\hat{\mathcal{A}}_{jilk}}{\mu_h},\quad \gamma=\frac{\hat{\mu}}{\mu_h},
\label{Nondimensionalization}
\end{aligned}
\end{equation}
where $\hat{\mu}$ is the uniform ground state shear modulus for the film and $\gamma$ measures the modulus ratio between the film and the substrate surface. To further simplify notations, we drop the asterisk from now on.

We obtain relationship between $U(y)$ and $V(y)$ from the linearized incompressibility condition \eqref{linearized-incompressibility}:
\begin{align}
    \dfrac{k \pi}{\lambda}U (y )+V^{'}(y )=0,
    \label{Stroh-incompressibility}
\end{align}
where a prime denotes a derivative with respect to $y$. Similarly, $P (y )$ can be solved from the expression of $\chi_{22}$:
\begin{align}
    P(y)=\dfrac{k \pi}{\lambda} \left(\mathcal{A} _{2211} -\mathcal{A} _{2222}-p  \right)U (y )-T _{22}(y ).
    \label{Stroh-stress}
\end{align}

Next, we define the displacement vector $\mathbf{u}(y )=[U (y ), V (y )]^{\textrm{T}}$, the traction vector $\mathbf{T}(y )=[T _{12}(y ), T _{22}(y )]^{\textrm{T}}$, and write their combination as a displacement-traction vector
\begin{align}
    \bm{\xi}(y )=[\mathbf{u}(y ), \mathbf{T}(y )]^{\textrm{T}}.
    \label{Displacement-traction}
\end{align}
Based on the equations \eqref{Incremental-equation}, \eqref{expansion}, \eqref{Stroh-incompressibility} and \eqref{Stroh-stress} we recast the boundary-value problem given by \eqref{Incremental-equation}-\eqref{sliding-end} as a  differential equation for the displacement-traction vector:
\begin{align}
     \bm{\xi}'(y )=\mathbf{Q}(y )\bm{\xi}(y ).
    \label{Equation-first-order}
\end{align}
This way of posing the problem is known as the $Stroh~formulation$ of the implied incremental problem \citep{Stroh1962JMP,Fu2005JE}, where $\mathbf{Q}(y )$ is the Stroh matrix possessing the following block formulation:
\begin{align}
    \mathbf{Q}(y )=\left[\begin{array}{cc}
        \mathbf{Q}_{1} & \mathbf{Q}_{2} \\
        \mathbf{Q}_{3} & \mathbf{Q}_{4}
        \end{array}\right],
    \label{Tensor-Q}
\end{align}
and $\mathbf{Q}_{i}~(i=1, 2, 3, 4)$ are $2 \times 2$ sub-blocks such that $\mathbf{Q}_{2}$ and $\mathbf{Q}_{3}$ are real symmetric, and $\mathbf{Q}_{4}=-\mathbf{Q}_{1}^\mathrm{T}$. In particular, the matrices $\mathbf{Q}_{1}$, $\mathbf{Q}_{2}$ and $\mathbf{Q}_{3}$ are given by
\begin{equation}
    \begin{aligned}
        &\mathbf{Q}_{1}=\dfrac{k \pi}{\lambda}\left[\begin{array}{cc}
            0 & \dfrac{\mathcal{A} _{2112}+p }{\mathcal{A} _{2121}} \\
            -1 & 0
        \end{array}\right], \quad
        \mathbf{Q}_{2}=\left[\begin{array}{cc}
            \dfrac{1}{\mathcal{A} _{2121}} & 0 \\
            0 & 0
        \end{array}\right], \\
        &\mathbf{Q}_{3}=\left(\dfrac{k \pi}{\lambda}\right)^{2}\left[\begin{array}{cc}
            \mathcal{A} _{1111}-\mathcal{A} _{2211}+\mathcal{A} _{2222}+2p  & 0 \\
            0 & \mathcal{A} _{1212}-\dfrac{\left(\mathcal{A} _{1221}+p \right)^2}{\mathcal{A} _{2121}}
        \end{array}\right].
        \label{Blocks}
    \end{aligned}
    \end{equation}
Note that in obtaining the expression of $\mathbf{Q}_{3}$ we have used the fact that $\mathcal{A} _{2112}=\mathcal{A} _{1221}$.

We apply the $impedance~matrix~method$ \citep{Biryukov1985,Biryukov1995,Shuvalov2003PRSA,Shuvalov2003QJMAM} to solve the incremental problem. To this end, we first assume a relation between the displacement vector $\mathbf{u}(y )$ and traction vector $\mathbf{T}(y )$:
\begin{align}
    \mathbf{u}(y )=\mathbf{K}(y )\mathbf{T}(y ),
        \label{Impedance}
\end{align} 
with $\mathbf{K}(y )$ an unknown matrix. It can be derived from the decay displacement boundary condition $\mathbf{u}(\infty)=\mathbf{0}$ that $\mathbf{K}(\infty)=\mathbf{0}$. Now substituting equations \eqref{Displacement-traction} and \eqref{Impedance} into \eqref{Equation-first-order} and eliminating the dependence of $\mathbf{T}(y )$, we obtain a $Riccati~equation$ for $\mathbf{K}$:
\begin{align}
    \mathbf{K}^{'}=\mathbf{Q}_{2}-\mathbf{K}\mathbf{Q}_{4}-\mathbf{K}\mathbf{Q}_{3}\mathbf{K}+\mathbf{Q}_{1}\mathbf{K},\quad h <y <\infty.
    \label{Riccati-substrate}
\end{align} 

For the film with a free lateral boundary, instead of equation \eqref{Impedance}, we use as an ansatz  a different relation between $\hat{\mathbf{u}}(y )$ and $\hat{\mathbf{T}}(y )$, which is written as
\begin{align}
    \hat{\mathbf{T}}(y )=\mathbf{Z}(y )\hat{\mathbf{u}}(y ),
    \label{Impedance-film}
\end{align} 
where $\mathbf{Z}(y )$ is known as the \textit{surface impedance matrix}. It follows  from the traction-free boundary condition $\hat{\mathbf{T}}(0)=\mathbf{0}$ that $\mathbf{Z}(0)=\mathbf{0}$. 

For the substrate, we obtain the following differential matrix \textit{Riccati~equation} by applying a similar derivation procedure:
\begin{align}
    \mathbf{Z}^{'}=\hat{\mathbf{Q}}_{3}-\mathbf{Z}\hat{\mathbf{Q}}_{1}-\mathbf{Z}\hat{\mathbf{Q}}_{2}\mathbf{Z}+\hat{\mathbf{Q}}_{4}\mathbf{Z},\quad 0<y <h .
    \label{Riccati-film}
\end{align}
The expressions of $\hat{\mathbf{Q}}_{i}$ $(i=1,2,3,4)$ are determined from \eqref{Blocks} by suitable variable substitutions.

We can solve numerically the equation \eqref{Riccati-substrate} subject to the boundary condition $\mathbf{K}(\infty)=\mathbf{0}$. In this way, the traction on the interface is given by $\mathbf{T}(h )=\mathbf{K}^{-1}(h )\mathbf{u}(h )$. Meanwhile, a similar solution procedure for equation \eqref{Riccati-film} can be applied and as a result the traction on the interface is written as $\mathbf{Z}(h )\hat{\mathbf{u}}(h )$. The continuity condition gives $\left(\mathbf{Z}(h )-\mathbf{K}^{-1}(h )\right)\mathbf{u}(h )=\mathbf{0}$. The existence of a non-trivial solution of $\mathbf{u}(h )$ then requires
\begin{align}
    \det \left( \mathbf{Z}(h )\mathbf{K}(h )-\mathbf{I}\right)=0,
    \label{Bifurcation-exact}
\end{align}
which is the bifurcation condition. This bifurcation condition \eqref{Bifurcation-exact} is valid for both problems considered in this study. Once the strain-energy function and the shear modulus distribution function are specified, one can obtain out the onset of surface wrinkling by solving \eqref{Bifurcation-exact} numerically.


\subsection{Preliminary for asymptotic analysis}\label{Preliminary}

In this subsection, we return  to the original incremental equations and transform it into an equivalent fourth-order ordinary differential equation suitable for asymptotic analysis. For that purpose, we substitute \eqref{expansion}, \eqref{Solution-form} into the governing system \eqref{linearized-incompressibility}, \eqref{Incremental-equation} and eliminate all unknowns except for $V(y)$. Ultimately, we acquire a dimensionless fourth-order ODE for $V (y )$ by utilizing the dimensionless variables in \eqref{Nondimensionalization}, which takes the following form
\begin{align}
    &V^{''''}+\frac{2 \mathcal{A}^{'}_{2121}}{\mathcal{A} _{2121}} V^{'''} +\frac{k^2\pi^2}{\lambda^{2} \mathcal{A} _{2121}}\left(\mathcal{A} _{1122}-\mathcal{A} _{1111}+2\mathcal{A} _{1221}+\mathcal{A} _{2211}-\mathcal{A} _{2222}\right)+\frac{\mathcal{A}^{''}_{2121}}{\mathcal{A} _{2121}} V^{''}   \notag\\&+\frac{k^2\pi^2}{\lambda^{2} \mathcal{A} _{2121}} \left(\mathcal{A}^{'}_{1122}-\mathcal{A}^{'}_{1111} +2 \mathcal{A}^{'}_{1221} +\mathcal{A}^{'}_{2211} -\mathcal{A}^{'}_{2222}\right)V^{'}  \notag \\
    &+\frac{k^2\pi^2}{\lambda^{2}\mathcal{A} _{2121}}\left(\frac{k^2 \pi^2}{\lambda^{2}}\mathcal{A} _{1212}+p^{''}+\mathcal{A}^{''}_{2112}\right) V=0.
    \label{Fourth-order}
\end{align}
Moreover, the traction components are given by
\begin{align}
T_{22}=&~\mathcal{A} _{2121}V^{'''} +\mathcal{A}^{'}_{2121}V^{''}-\dfrac{ k^2 \pi^2}{\lambda^2}\left(p +\mathcal{A} _{1111}-2\mathcal{A} _{1122}-\mathcal{A} _{2112}+\mathcal{A} _{2222}\right)V^{'} \notag \\
    &+\dfrac{ k^2 \pi^2}{\lambda^2}\left(p^{'}+\mathcal{A}^{'}_{2112}\right)V,\label{Traction-component1}\\
T_{12}=&~\mathcal{A} _{2121}V^{''}+ \dfrac{ k^2 \pi^2}{\lambda^2}\left(p +\mathcal{A} _{2112} \right)V,
\label{Traction-component2}
\end{align}
and the boundary conditions and continuity conditions \eqref{Incremental-bc}-\eqref{Incremental-dc} are transformed into
\begin{equation}
\begin{aligned}
    &\hat{T}_{12}=0, \quad \hat{T}_{22}=0, \quad \textrm{on}~y =0,\\
    &\hat{T}_{12}=T_{12}, \quad \hat{T}_{22}=T_{22}, \quad \hat{V}=V,\quad \hat{V}^{'}=V^{'}, \quad \textrm{on}~y =h,\\
    &V=0,\quad V^{'}=0, \quad \textrm{as}~y \rightarrow \infty.
    \label{BC}
\end{aligned}
\end{equation}

This formulation is valid for both the compression and the growth cases.


\section{WKB reduction}\label{WKB}
We consider the short wavelength approximation, i.e. $k\rightarrow \infty$, to carry out an asymptotic analysis on the critical condition of surface wrinkling, following \citep{Jin2018IJES}.

First, to make progress we must choose a material model. Here, we will use the neo-Hookean model but other choices can be used within the same setting. Note that an asymptotic formula for the critical buckling stretch has been derived by \cite{Cai1999PRSA} when the substrate is homogeneous and the incompressible neo-Hookean model is employed. This will be used as a benchmark for validating the asymptotic solutions in our particular situation.
In the plane-strain case, the strain-energy functions for the film and substrate are given by
\begin{align}
    \hat{W}=\frac{1}{2} \hat{\mu}\left(I_1-2\right), \quad
    W=\frac{1}{2}\mu_\textrm{s}(X_2)\left(I_1-2\right),
    \label{neo-Hookean}
\end{align}
where $\hat{\mu}$ is the homogeneous ground state shear modulus of the film, $\mu_\textrm{s}(X_2)$ is the modulus function, and $I_1=\operatorname{tr} \mathbf{F}\mathbf{F}^\textrm{T}$ if the external load is given by axial compression or $I_1=\operatorname{tr} \mathbf{A}\mathbf{A}^\textrm{T}$ in the growth case. 
Before proceeding further, we introduce the dimensionless modulus 
\begin{align} 
    \mu(y)=\dfrac{\mu_\textrm{s}}{\mu_{h}},
    \label{ratio}
\end{align}
where $\mu_{h}$ has been defined in \eqref{Nondimensionalization}.


\subsection{Uniaxial compression}\label{Uniaxial-compression}
For the neo-Hookean model \eqref{neo-Hookean}, the fourth-order ODE \eqref{Fourth-order}  simplifies to 
\begin{align}
    V^{''''}+\frac{2 \mu^{'}}{\mu} V^{'''}+\left(\frac{\mu^{''}}{\mu  }- \frac{k^2\pi^2\left(\lambda ^4+1 \right)}{\lambda^{2}}\right) V^{''}-\frac{k^2\pi^2\left(\lambda ^4+1 \right) \mu^{'}}{\lambda^{2} \mu} V^{'} +k^2\pi^2\left(\frac{\mu^{''}}{\lambda^{2} \mu}+k^2\pi^2\right) V=0.
    \label{ODE-compression}
\end{align}
The boundary conditions and continuity conditions in \eqref{BC} become
\begin{equation}
\begin{aligned}
    &\hat{V}^{'''} -\frac{k^2\pi^2\left(\lambda ^4+2\right)}{\lambda^{2}}\hat{V}^{'}=0, \quad \hat{V}^{''}+ \dfrac{k^2 \pi^2}{\lambda^2}\hat{V}=0, \quad \textrm{on}~y =0,\\
    &\gamma\hat{V}^{'''}-\mu V^{'''}-\mu^{'}V^{''}-\frac{\gamma k^2\pi^2\left(\lambda ^4+2\right)}{\lambda^{2}}\hat{V}^{'}+\frac{k^2\pi^2\left(\lambda ^4+2\right) \mu}{\lambda^{2}} V^{'}-\frac{k^2\pi^2\mu^{'}}{\lambda^{2}}=0, \quad \textrm{on}~y =h ,\\
    &\gamma\hat{V}^{''}-\mu V^{''}+ \gamma \dfrac{k^2 \pi^2}{\lambda^2}\hat{V}-\frac{k^2\pi^2\mu }{\lambda^{2}}V=0, \quad \textrm{on}~y =h ,\\
    &\hat{V}-V=0,
\quad\hat{V}^{'}-V^{'}=0, \quad \textrm{on}~y =h ,\\
    &V=0,\quad V^{'}=0, \quad \textrm{as}~y \rightarrow \infty.
    \label{BC-compression}
\end{aligned}
\end{equation}

The above equation \eqref{ODE-compression} is for the substrate; its counterpart for the film can be found by replacing $V$ by $\hat{V}$ and by setting all derivative terms of $\mu $ to zero as the film is composed of a homogeneous material. Accordingly, the fourth-order ODE for the film is given by
\begin{align}
    \hat{V}^{''''}-\frac{k^2\pi^2\left(\lambda ^4+1\right)}{\lambda^{2}} \hat{V}^{''}+ k^4 \pi^4 \hat{V}=0.
    \label{Strain-energy}
\end{align}
Solving directly the associated characteristic equation yields four eigenvalues $s_{1,2}=\mp k \pi/\lambda$ and $s_{3,4}=\mp  k \pi \lambda$.
Correspondingly, the general solution to equation \eqref{Strain-energy} can be written as
\begin{align}
    \hat{V} (y )=C_1 \textrm{exp}(s_{1}y )+C_2 \textrm{exp}(s_{2}y )+C_3 \textrm{exp}(s_{3}y )+C_4 \textrm{exp}(s_{4}y ),
    \label{Solution-Vhatstar}
\end{align}
where $C_{i}~(i=1,2,3,4)$ are constants.

Following the WKB method,  we seek for a solution of \eqref{ODE-compression} by using the following ansatz \citep{Hinch1991}
\begin{align}
    V (y )=\textrm{exp}\left(\int^{\infty}_{y } S(\tau) \textrm{d}\tau\right).
    \label{WKB-form}
\end{align}
Substituting the solution \eqref{WKB-form} into  \eqref{ODE-compression} gives
\begin{align}
    &S^4-\frac{2 \mu^{'}}{\mu} \left(S^3+S''-3SS'\right)-6 S^2 S'+4 SS''+3 S'^2-S^{'''} +\left(\frac{\mu^{''}}{\mu}-\frac{k^2\pi^2\left(\lambda ^4+1\right)}{\lambda^{2}}\right) \left(S^2-S'\right) \notag \\
    &+\frac{k^2\pi^2\left(\lambda ^4+1\right) \mu^{'}}{\lambda^{2} \mu} S+k^4\pi^4+\frac{k^2\pi^2\mu^{''}}{\lambda ^2\mu}=0.
    \label{WKB-equation}
\end{align}

The next step is to look for an asymptotic solution to the above equation of the form
\begin{align}
    S(y )=k S_{0}+S_{1}+k^{-1}S_{2}+ \cdots=\sum\limits^{\infty}\limits_{m=0}k^{1-m}S_{m},
    \label{WKB-expansion}
\end{align}
where $S_{m}~(m=0,1,2\dots)$ are unknown functions of $y $. Substituting \eqref{WKB-expansion} into  \eqref{WKB-equation} and equating all coefficients of like powers of $1/k$ to zero provide infinitely many equations. In particular, the leading-order equation is a quartic algebraic one of $S_0$ from which four independent solutions can be obtained. Typically, a two-term approximation containing $S_0$ and $S_1$ in \eqref{WKB-expansion} already provides sufficient accuracy \citep{Jin2018IJES}. The four independent solutions for $S(y)$ are:
\begin{equation}
    S^{(1,2)}(y )=\mp\dfrac{k \pi}{\lambda}+\frac{\mu^{'}}{2 \mu}+\cdots,\quad 
    S^{(3,4)}(y )=\mp k \pi \lambda+\frac{\mu^{'}}{2 \mu}+\cdots.
    \label{Foursol}
\end{equation}
Then, the general solution for $V (y )$ can be written as
\begin{align}
    V (y )=\sum^{4}_{i=1}C_{i+4}\textrm{exp}\left(\int^{\infty}_{y }S^{(i)}(\tau) \textrm{d}\tau  \right),
    \label{Solution-Vstar}
\end{align}
where $C_{i+4}$ $(i=1,2,3,4)$ are arbitrary constants. Inserting \eqref{Solution-Vhatstar} and \eqref{Solution-Vstar} into \eqref{BC-compression}, we obtain a vector equation
\begin{align}
    \mathbf{M}\mathbf{C}=\bm 0,
    \label{Matrix}
\end{align}
where $\mathbf C=[C_1,C_2,C_3,C_4,C_5,C_6,C_7,C_8]^\textrm{T}$ and the coefficient matrix $\mathbf{M}$ is given by
\begin{align}
    &\mathbf{M}={\left[
        \begin{array}{cccccccc}
            \hat{a}_{1} & \hat{a}_{2} & \hat{a}_{3} & \hat{a}_{4} & 0 & 0 & 0 & 0\\
            \hat{b}_{1} & \hat{b}_{2} & \hat{b}_{3} & \hat{b}_{4} & 0 & 0 & 0 & 0\\
            q_{1}\hat{a}_{1} & q_{2}\hat{a}_{2} & q_{3}\hat{a}_{3} & q_{4}\hat{a}_{4} & f_{1}a_{1} & f_{2}a_{2} & f_{3}a_{3} & f_{4}a_{4}\\
            q_{1}\hat{b}_{1} & q_{2}\hat{b}_{2} & q_{3}\hat{b}_{3} & q_{4}\hat{b}_{4} & f_{1}b_{1} & f_{2}b_{2} & f_{3}b_{3} &f_{4}b_{4}\\
            q_{1} & q_{2} & q_{3} & q_{4} & -f_{1} & -f_{2} & -f_{3} & -f_{4}\\
            q_{1}s_{1} & q_{2}s_{2} & q_{3}s_{3} & q_{4}s_{4} & f_{1}S^{(1)}(h ) & f_{2}S^{(2)}(h ) & f_{3}S^{(3)}(h ) & f_{4}S^{(4)}(h )\\
            0 & 0 & 0 & 0 & 1 & 1 & 1 & 1\\
            0 & 0 & 0 & 0 & -S^{(1)}(\infty) & -S^{(2)}(\infty) & -S^{(3)}(\infty) & -S^{(4)}(\infty)\\
        \end{array}
    \right] }.    \label{Expression-M}
\end{align}
Note that the parameters $\hat{a}_{i}, \hat{b}_{i}, a_{i}, b_{i}, q_{i},$ and $f_{i}$ are defined as
\begin{equation}
\begin{aligned}
    &\hat{a}_{i}=\gamma s_{i}\left(s_{i}^2-\frac{k^{2}\pi^{2}\left( \lambda ^4+2 \right)}{\lambda^2}\right),\quad
    \hat{b}_{i}=\gamma\left(s_{i}^2+\frac{k^{2}\pi^{2}}{\lambda^2}\right),\quad
    q_{i}=\textrm{exp}(s_{i} h ),  \\
    &a_{i}= S^{(i)}(h )^3 \mu (h )-S^{(i)}(h )^2\mu^{'}(h )
     - S^{(i)}(h )\mu (h )\left(3S^{(i)'}(h )+\frac{k^{2}\pi^{2}}{\lambda^2}\left(\lambda ^4+2\right)\right), \\
    &~~~~~~~+ S^{(i)''}(h )\mu (h )+\left(S^{(i)'}(h )-\frac{k^{2}\pi^{2}}{\lambda^2}\right)\mu^{'}(h ) , \\
    &b_{i}=\mu  (h )\left(S^{(i)'}(h )-S^{(i)}(h )^2-\frac{k^{2}\pi^{2}}{\lambda^2}\right),\quad 
    f_{i}=\textrm{exp}\left(\int^{\infty}_{h }S^{(i)}(\tau) \textrm{d}\tau \right).
    \label{Coefficients}
\end{aligned}
\end{equation}

The existence of a non-trivial solution to \eqref{Matrix} requests the vanishing of the determinant of $\mathbf{M}$. Finally, the bifurcation condition reads
\begin{align}
    \det\mathbf{M}=0,
    \label{Determinant}
\end{align}
which can be regarded as a function of the principal stretch $\lambda$, the wavenumber $k$, the modulus ratio $\gamma$, the thickness $h_0 $ and the parameters involved in the modulus function $\mu (y )$. After a careful observation of the coefficients in \eqref{Coefficients}, we find that $f_{2}$ and $f_{4}$ are both exponentially large so the terms whose factor is $f_2f_4$ will dominate the value of determinant. As a result, the bifurcation condition \eqref{Determinant} can be rewritten as
\begin{align}
    \Xi=\frac{\det\mathbf{M}}{f_{2} f_{4}} - E.S.T.=0,
    \label{f}
\end{align}
where $E.S.T$ denotes exponential small terms which will be neglected in further analysis and the algebraic equation $\Xi=0$ provides a  way to obtain an analytical formula for the critical bifurcation threshold.


\subsection{Restricted growth}\label{Restricted-growth}
When the compressive stress is induced by growth, the bifurcation parameter changes is the growth factor. The strain-energy functions for the film and the substrate are provided by \eqref{neo-Hookean}. In this case, the dimensionless fourth-order ODE \eqref{Fourth-order} is
\begin{align}
    &V^{''''}(y )+\frac{2 \mu^{'}}{\mu} V^{''''}+\left(\frac{\mu^{''}}{\mu }-k^2\pi^2\left(\frac{1}{g_1^4}+1\right)\right) V^{''}-\frac{k^2\pi^2\left(g_1^4+1\right) \mu^{'}}{g_1^4 \mu} V^{'}\notag\\&+\left(\frac{k^2\pi^2\mu^{''}}{\mu}+\frac{k^4 \pi^4}{g_1^4}\right) V =0,
    \label{ODE-growth}
\end{align}
where $g_1$ is the growth factor for the substrate in the horizontal direction. Correspondingly, the incremental boundary conditions and continuity conditions in \eqref{BC} can be written as
\begin{equation}
    \begin{aligned}
        &\hat{V}^{'''} -k^2\pi^2\left(\frac{1}{\hat{g}_1^4}+2\right) \hat{V}^{'}=0, \quad \hat{V}^{''}+ k^2\pi^2\hat{V}=0, \quad \textrm{on}~y =0,\\
        &\gamma \hat{g}_1^2 \hat{V}^{'''}-g_1^2 \mu V^{'''}-{g}_1^2 \mu^{'} V^{''}-\frac{\gamma k^2\pi^2\left(2 \hat{g}_1^4 +1 \right)}{\hat{g}_1^2}\hat{V}^{'}+\frac{ \mu k^2\pi^2\left(2 {g}_1^4 +1 \right)}{{g}_1^2}V^{'}-g_1^2 k^2 \pi^2\mu^{'}V=0, \quad \textrm{on}~y =h ,\\
        &\gamma \hat{g}_1^2 \hat{V}^{''}-{g}_1^2 \mu V^{''}+ \gamma \hat{g}_1^2 k^2 \pi^2\hat{V}-{g}_1^2 k^2 \pi^2\mu V=0, \quad \textrm{on}~y =h ,\\
        &\hat{V}-V=0, \quad \hat{V}^{'}-V^{'}=0, \quad \textrm{on}~y =h ,\\
        &V=0,\quad V^{'}=0, \quad \textrm{as}~y \rightarrow \infty.
        \label{BC-growth}
    \end{aligned}
    \end{equation}
In our previous study \citep{Liu2021PRSA}, it was found that a gradient in the growth function has a negligible effect on the critical buckling threshold and a constant growth can provide a good approximation on interpreting complicated morphological formations in bilayer systems. Thus, we assume that both layers share the same constant growth factor such that $\hat{g}_1=\hat{g}_2=g_1=g_2=g$. Next, we aim to  solve asymptotically the eigenvalue problem with variable coefficients arising from \eqref{ODE-growth} and \eqref{BC-growth}.

The fourth-order ODE for the homogeneous film can be obtained by removing all derivative terms of the shear modulus in \eqref{ODE-growth}. Solving directly the associated characteristic equation gives a general solution of  $\hat{V}(y)$ of the form
\begin{align}
    \hat{V}(y)=C_1 \textrm{exp}(r_{1}y )+C_2 \textrm{exp}(r_{2}y )+C_3 \textrm{exp}(r_{3}y )+C_4 \textrm{exp}(r_{4}y ), 
    \label{eq5.1}
\end{align}
where $C_i$ $(i=1,2,3,4)$ are constants and the eigenvalues $r_{i}$ $(i=1,2,3,4)$ are given by $r_{1,2}=\mp k\pi$ and $r_{3,4}=\mp k \pi/g^2$.

To solve \eqref{ODE-growth}, we use again a WKB-type method with ansatz  given by
\begin{align}
    V (y )=\textrm{exp}\left(\int^{\infty}_{y } R(\tau) \textrm{d}\tau\right),
    \label{Expansion-RR}
\end{align}
and the integrand function can be expanded in $k$ as
\begin{align}
   R(y )=k R_{0}+R_{1}+k^{-1}R_{2}+\cdots=\sum\limits^{\infty}\limits_{m=0}k^{1-m}R_{m},  
    \label{Expansion-R}
\end{align}
where $R_{m}$ $(m=0,1,2\dots)$ are unknown functions of $y $. On substituting equations \eqref{Expansion-RR} and \eqref{Expansion-R} into \eqref{ODE-growth} we also figure out four solutions of $R_0$ from the leading-order equation. Similar to the solution procedure illustrated in Section \ref{Uniaxial-compression}, we are able to obtain the following four independent solutions
\begin{align}
    R^{(1,2)}(y )=\mp k \pi+\dfrac{\mu^{'}(y )}{2 \mu  (y )}+\cdots,\quad 
    &R^{(3,4)}(y )=\mp\dfrac{k \pi}{g^2}+\dfrac{\mu^{'}(y )}{2 \mu  (y )}+\cdots,
    \label{Foursol-growth}
\end{align}
The general form of $V (y )$ is
\begin{align}
    V (y )=\sum^{4}_{i=1}C_{i+4}\textrm{exp}\left(\int^{\infty}_{y }R^{(i)}(\tau) \textrm{d}\tau\right),
    \label{Solution-Vstar-growth}
\end{align}
where $C_{i+4}$ $i=1,2,3,4$ are constants. We further substitute \eqref{eq5.1} and \eqref{Solution-Vstar-growth} into \eqref{BC-growth} to obtain a vector equation of the same form as \eqref{Matrix} but with all $S^{(i)}$ and $s_i$ in \eqref{Expression-M} replaced by $R^{(i)}$ and $r_i$, and entries given by
\begin{equation}
\begin{aligned}
    &\hat{a}_{i}=\gamma r_{i}\left(g^4 r_{i}^2-k^2 \pi^2 \left(2 g^4 +1 \right)\right),\quad
    \hat{b}_{i}=\gamma\left(r_{i}^2+k^{2}\pi^{2}\right),\quad
    q_{i}=\textrm{exp}(r_{i} h ), \\
    &a_{i}= g^4 \left( R^{(i)}(h )^3 \mu (h )- R^{(i)}(h )^2\mu^{'}(h )R^{(i)''}(h )\mu (h )+ \left(R^{(i)'}(h )-k^2 \pi^2 \right) \right), \\
    &~~~~~~~- R^{(i)}(h )\mu (h )\left(3 g^4 R^{(i)'}(h )+k^2 \pi^2 \left( 2 g^4 +1\right)\right),\\
    &b_{i}=\mu  (h )\left(R^{(i)'}(h )-R^{(i)}(h )^2-k^{2}\pi^{2}\right),\quad 
    f_{i}=\textrm{exp}\left(\int^{\infty}_{h }R^{(i)}(y ) \textrm{d}y  \right).
\end{aligned}
\end{equation}

As in the previous case, the WKB analysis  generates an approximate bifurcation condition $\bar{\Xi}=0$ from $\det\bar{\mathbf{M}}=0$, which will be used later on.

\subsection{The shear modulus various exponentially}

The problem cannot be solved for an arbitrary modulus function. Hence, we consider a modulus with an exponential decay of the form
\begin{align}
    \mu_\textrm{s}(X_2)=\mu_0 \textrm{exp}\left(-\zeta (X_2-h_0)\right)+\mu_1, \quad X_2\geqslant h_0,\quad\textrm{the~original~notations~with~dimension}
    \label{Modulus-exponential}
\end{align}
where $\mu_0$, $\mu_1$ are two constants, and $\zeta>0$ measures how fast the modulus decays ($\zeta>0$ and $(\mu_0+\mu_1)/\mu_1>1$) or grows (either $\zeta<0$ or $\zeta>0$ and $(\mu_0+\mu_1)/\mu_1<1$ ) from the surface of substrate. If $\zeta>0$, we find $\mu_0+\mu_1=\mu(h_0)=\mu_h$ and $\mu_1=\mu(\infty)=\mu_\infty$, or otherwise we require $\mu_0+\mu_1=\mu_h$ and $\mu_\infty=\infty$.  Then
the dimensionless form of \eqref{Modulus-exponential} with $y $ as a variable reads
\begin{align}
    \mu (y )=\left(1-\beta \right) \textrm{exp}\left(-\alpha  (\lambda y -h _{0})\right)+\beta, \quad y \geqslant h ,
    \label{Modulus-exponential-N}
\end{align}
where $\beta=\mu_\infty/\mu_h\leqslant1$ and $\alpha=\zeta L$ is the normalized decay rate. We also impose $\mu_1\neq0$ to avoid zero modulus \cite{Chen2017PRSA}. 
Using  \eqref{Modulus-exponential-N}, we are able to express the bifurcation condition \eqref{f} as $\Xi(\lambda, k, \alpha, \beta, \gamma, h _{0})$. For the growth scenario, the exponential modulus function \eqref{Modulus-exponential-N} is used as well but we substitute $\lambda=1/g^2$ to obtain a condition in terms of of $g$ of the form  $\bar{\Xi}=\bar{\Xi}(g, k, \alpha, \beta, \gamma, h _{0})$. 

A particularly interesting situation arises when the substrate is more compliant that the thin film with $\gamma$  large and $h_0 $ is small considered in \cite{Cao2012IJSS} and \cite{Chen2017PRSA}. Our study unifies and generalises both work by considering the two cases $\beta\sim\mathcal{O}(1)$ and large $\beta$.


\section{Asymptotic solution when $\beta$ is small or of $\mathcal{O}(1)$}\label{Asymptotic-analysis-decay}
We focus on $\beta\ll1$ or $\beta\sim\mathcal{O}(1)$ and  derive some asymptotic estimates for the critical buckling load as well as for the critical wavenumber. The  compression and growth cases will be explored separately. 


\subsection{Uniaxial compression}\label{Uniaxial-compression1}

\begin{figure}[!htp]
    \centering\includegraphics[scale=1]{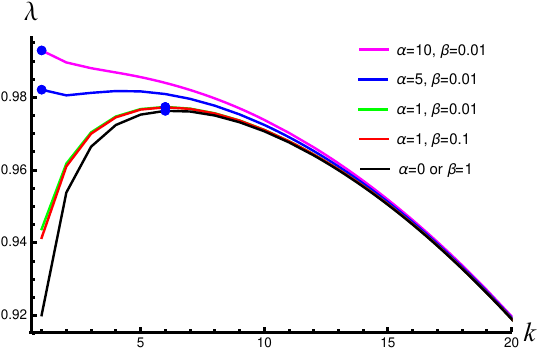}
    \caption{Bifurcation curves based on the exact bifurcation condition \eqref{Bifurcation-exact} for $\gamma=100,~h _{0}=0.015$. The blue dots determine the critical principal stretch $\lambda_{\textrm{cr}}$ associated with the critical wavenumber $k_{\textrm{cr}}$. }
    \label{Lambda-mode}
\end{figure}

Before carrying out the asymptotic analysis, we first present five bifurcation curves according to the exact bifurcation condition \eqref{Bifurcation-exact} in Figure \ref{Lambda-mode} when the modulus ratio $\gamma=100$ and the initial film thickness $h _{0}=0.015$. The purpose is to identify some valid parameter domains where the short wavelength assumption can be satisfied. In Figure \ref{Lambda-mode} we see that each curve has a maximum, highlighted by a blue point; its  vertical coordinate gives the critical stretch $\lambda_{\textrm{cr}}$ triggering surface wrinkling while the horizontal coordinate gives the critical wavenumber $k_{\textrm{cr}}$. When $\alpha=5$ or $\alpha=10$ the critical wavenumber $k_{\textrm{cr}}$ is small, which does not satisfy the short wavelength assumption in the WKB expansion \eqref{WKB-form}. Hence, we consider the case where $\alpha\sim\mathcal{O}(1)$. We also note that the specific value of $\beta$ when $\beta\sim\mathcal{O}(1)$ has a weak effect on the critical state when the modulus ratio $\gamma$ is large.
\begin{figure}[!htp]
    \centering\includegraphics[scale=0.98]{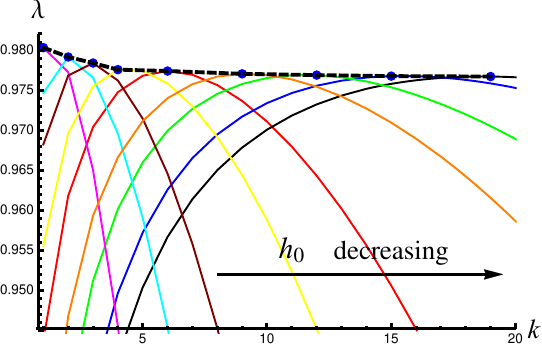}
    \caption{Bifurcation curves based on the exact bifurcation condition \eqref{Bifurcation-exact} for $\gamma=100,~\alpha=1,~\beta=0.01$ and $h _{0}=0.06,~0.04,~0.03,~0.02,~0.015,~0.01,~0.008,~0.006,~0.005$, respectively. Dashed line is the envelope of the bifurcation curves.}
    \label{Lambda-k-fixedgamma}
\end{figure}

Next, we explore the influence of the film initial thickness $h _{0}$ on the critical state. Given $\gamma=100,~\alpha=1,~\beta=0.01$, Figure \ref{Lambda-k-fixedgamma} shows nine bifurcation curves stemming from  \eqref{Bifurcation-exact} for different values of $h _{0}$. The dashed line is the envelope of the maximums of bifurcation curves. We  observe that the critical wavenumber $k_{\textrm{cr}}$ is a decreasing function of $h _{0}$. To satisfy the short wavelength assumption, we  confine our attention to small $h _{0}$ as found  in many engineering devices or biological tissues. It is worth noting that $\lambda_{\textrm{cr}}$ is independent of $h _{0}$ when the substrate is homogeneous \citep{Cai1999PRSA,Liu2014IJES,Alawiye2019PTRSA,Wang2023MMS}, but in our problem $\lambda_{\textrm{cr}}$ is no longer a constant as shown in Figure \ref{Lambda-k-fixedgamma}. In addition, we assume that the film is much stiffer than the substrate, which implies that $\gamma$ is large. It is also known that $\varepsilon=1-\lambda_{\textrm{cr}}$ is small as the film is much stiffer than the substrate.  In conclusion, we will place ourselves in the situation where the following four parameters are small: $1/k, h _{0}, 1/\gamma$ and $\varepsilon$.

 \begin{figure}[!htp]
    \centering\includegraphics[scale=1]{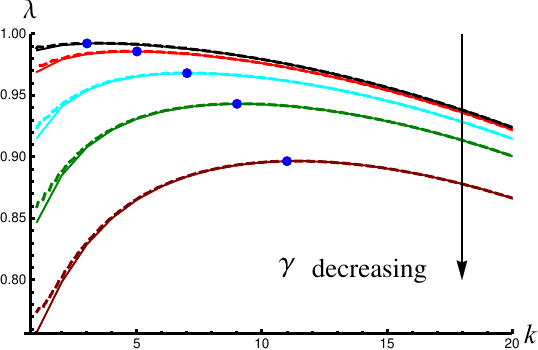}
    \caption{Comparisons of the bifurcation curves between the approximate (dashed line) and exact (solid line) bifurcation conditions when $\gamma=500,~200,~60,~25,~10$ and $h _{0}=0.015,\alpha=1,\beta=0.01$. The blue dots highlight the maximum that corresponds to the critical principal stretch $\lambda_{\textrm{cr}}$ associated with the critical wavenumber $k_{\textrm{cr}}$.}
    \label{Lambda-k-fixedh0}
\end{figure}

In Figure \ref{Lambda-k-fixedh0}, we compare the bifurcation curves based on the approximate bifurcation condition \eqref{f} (dashed line) and the exact one \eqref{Bifurcation-exact} (solid line) if $10 \leqslant \gamma \leqslant 500$ and $h _{0}=0.015,\alpha=1,\beta=0.01$. It can be seen that the asymptotic result is in good agreement  with the exact (numerical) one for all cases, even when $k_\textrm{cr}=3$. Similarly, we checked that the asymptotic bifurcation condition \eqref{f} remains valid when $\alpha\sim\mathcal{O}(1)$, $1/k, h _{0}, \varepsilon$, and $1/\gamma$ are all small parameters. Therefore, we can now start from this approximation \eqref{f} to obtain explicit solution for $\lambda_{\textrm{cr}}$ and $k_{\textrm{cr}}$.
 

\subsubsection{Very stiff film on soft graded substrate: $\gamma\sim\mathcal{O}(h_{0}^{-\frac{3}{2}})$}\label{case1}
First, we expand equation \eqref{f} in terms of the small parameters $\varepsilon$ and $h _{0}$ to obtain
\begin{align}
    &\varepsilon ^2 \bigg(t_{0}k^{4}+t_{1}k^{3} +t_{2} k^5  h _0+t_{3} k^4 h _0 +t_{4} k^6 h^{2}_{0}+t_{5} k^5 h_{0}^{2}   +\cdots + \gamma\left(t_{6} k^5  h _0 + t_{7} k^4 h _{0} \notag \right.\\
    &\left.+t_{8} k^6 h_{0}^{2}+t_{9} k^5 h_{0}^{2}+\cdots \right) + \gamma^{2}\left( t_{10}k^{8}h_{0}^{4}+ t_{11}k^{9}h_{0}^{5} \right)+\cdots \bigg) + \varepsilon ^3 \bigg( t_{12} k^{4}+t_{13} k^{3}\notag \\
    &+t_{14} k^5  h _0+t_{15} k^4 h _0 +t_{16} k^6 h_{0}^{2}+t_{17} k^5 h_{0}^{2} +\cdots + \gamma\left(t_{18} k^5  h _0 + t_{19} k^4 h _{0}+t_{20} k^6 h_{0}^{2}\notag \right.\\
    &\left.+t_{21} k^5 h_{0}^{2}+\cdots \right)+\gamma^{2}\left( t_{22}k^{6} h_{0}^{2}+t_{23}k^{7} h_{0}^{3}+t_{24}k^{8} h_{0}^{4} \right)+ \cdots\bigg)=0,
    \label{Asymptotic-expansion-case1}
\end{align}
where $t_{i}$ $(i=0,1,2,\dots)$ are constant coefficients related to $\alpha$ and $\beta$, whose lengthy expressions are not written out for brevity. The leading-order balance and principle of least degeneracy \citep{Liu2014IJES} are used to derive the order relations among $\varepsilon$ and $k h _{0}$ and $\gamma$. Keeping all possible leading-order terms,  \eqref{Asymptotic-expansion-case1} can be written as
\begin{align}
    t_{0}+ \gamma\left(t_{6} k  h _0+t_{8}k^2 h_{0}^{2}\right)+ \gamma^{2}\left(t_{10}k^{4}h_{0}^{4}+ t_{11}k^{5}h_{0}^{5} \right)+\varepsilon \gamma^{2} \left( t_{22}k^{2} h_{0}^{2}+t_{23}k^{3} h_{0}^{3} \right)+I.H.O.=0,
    \label{Leading-order}
\end{align}
where $t_{0}\sim\mathcal{O}(1)$ and $I.H.O$ represents infinitesimal of higher order. Analyzing and discussing all possible balances following \citep{Jin2018IJES,Liu2014IJES}, we find that when $\varepsilon \sim\mathcal{O}(\gamma ^{-\frac{2}{3}})$ and $k h _{0}\sim\mathcal{O}(\gamma ^{-\frac{1}{3}})$, the leading terms in  \eqref{Asymptotic-expansion-case1} are of $\mathcal{O}(\gamma ^{\frac{2}{3}})$ and  given by $t_{6}kh _{0}\gamma$, $t_{10}k^{4}h_{0}^{4}\gamma^{2}$ and $t_{22}k^{2} h_{0}^{2}\varepsilon \gamma^{2}$. Then, we extract all possible second-order terms for equation \eqref{Asymptotic-expansion-case1} and write
\begin{align}
    \mathcal{O}(\gamma ^{\frac{2}{3}})+ t_{0} +\gamma\left(t_{7}h _{0}+t_{8}k^2 h_{0}^{2}\right)+\gamma^{2}\left(t_{11}k^{5}h_{0}^{5}+t_{23}k^{3} h_{0}^{3} \varepsilon\right)+I.H.O.=0,
    \label{Second-order}
\end{align}
where the terms $t_{8}k^2 h_{0}^{2}\gamma$, $t_{11}k^{5}h_{0}^{5}\gamma^{2}$ and $t_{23}k^{3} h_{0}^{3} \varepsilon \gamma^{2}$ are all of order $\mathcal{O}(\gamma ^{\frac{1}{3}})$. Using the assumption that $\gamma$ is a large parameter, we  find that $\mathcal{O}(\gamma ^{\frac{1}{3}}) \gg t_{0}$ and the second-order must be equal to $\mathcal{O}(\gamma ^{\frac{1}{3}})$. According to the least degeneracy principle, in order to keep the $t_{7} h _0 \gamma$ term, we can determine $h _{0}\sim\mathcal{O}(\gamma ^{-\frac{2}{3}})$ and $k\sim\mathcal{O}(\gamma ^{\frac{1}{3}})$. Through the above analysis, the order relations are clearly given by $\varepsilon\sim\mathcal{O}(\gamma ^{-\frac{2}{3}})$, $h _{0}\sim\mathcal{O}(\gamma ^{-\frac{2}{3}})$ and $k\sim\mathcal{O}(\gamma ^{\frac{1}{3}})$.

To derive an asymptotic solution for $\lambda_{\textrm{cr}}$ and $k_{\textrm{cr}}$, in addition to the equation \eqref{Asymptotic-expansion-case1} we  also need to compute $\partial \lambda/\partial k=0$, or equivalently
\begin{align}
\frac{\partial \Xi}{\partial k}=0.
    \label{Dfdk}
\end{align}
Making use of the order relations we use
\begin{align}
    h _{0}=a_{0} \gamma ^{-\frac{2}{3}},
    \label{h0}
\end{align}
where $a_{0}$ is a constant of $\mathcal{O}(1)$. We then seek an asymptotic solution for $\lambda_{\textrm{cr}}$ and $k_{\textrm{cr}}$ in the form of 
\begin{equation}
\begin{aligned}
 &\lambda_{\textrm{cr}} =1-\varepsilon=1-m_{0}\gamma ^{-\frac{2}{3}} -m_{1}\gamma^{-1} -m_{2}\gamma ^{-\frac{4}{3}} -m_{3}\gamma ^{-\frac{5}{3}}+\dots, \\
 &k_{\textrm{cr}}=l_{0} \gamma ^{1/3}+ l_{1}+l_{2}\gamma ^{-\frac{1}{3}} +l_{3}\gamma ^{-\frac{2}{3}} \dots,
\end{aligned} \label{Asymptotic-form}
\end{equation}
where $m_{i}$ and $l_{i}~(i=0,1,2\dots)$ are coefficients to be determined.

Substituting \eqref{Asymptotic-form} into \eqref{f} and \eqref{Dfdk} provides the four-term asymptotic solutions
\begin{align}
    \lambda _{\textrm{cr}}=&~1-\frac{3^{2/3}}{4 \gamma ^{2/3}}-\frac{3^{1/3} \alpha  \left(\beta -1\right)h _{0}}{4 \gamma^{1/3} }+\frac{1}{480} \left(\frac{99\ 3^{1/3}}{\gamma ^{4/3}}+\frac{40\ 3^{2/3} \alpha  \left(\beta -1\right) h _0}{\gamma ^{2/3}}+100 \alpha ^2 \left(\beta -1\right)^2 h_0^{2}\right) \notag\\
    &-\frac{1}{720} \bigg(\frac{90\ 3^{2/3}}{\gamma ^{5/3}}-\frac{189 \alpha  \left(\beta -1\right) h _0}{\gamma }+\frac{30\ 3^{1/3} \alpha ^2 \left(\beta ^2-3 \beta +2\right) h_0^{2}}{\gamma^{1/3}} \notag\\
    &+10\ 3^{2/3} \alpha ^3 \left(10 \beta ^3-32 \beta ^2+33 \beta -11\right) \gamma^{1/3} h_0^{3}\bigg)+\mathcal{O}(\gamma^{-2}),
    \label{Solution-lambda}
\end{align}
\begin{align}
    k_{\textrm{cr}}=&~\frac{3^{1/3}}{\pi \gamma^{1/3} h _{0}}+\frac{\alpha  \left(\beta -1\right)}{\pi} -\frac{1}{60\pi}\left(\frac{81}{\gamma  h _0}+\frac{10\ 3^{1/3} \alpha  \left(\beta -1\right)}{\gamma ^{1/3}}  +35\ 3^{2/3} \alpha ^2 \left(\beta -1\right)^2 \gamma ^{1/3} h _0 \right)\notag \\
    &+\frac{\alpha  \left(\beta -1\right)}{180\pi} \left(\frac{42\ 3^{2/3}}{\gamma ^{2/3}}+ 30 \alpha  \left(2 \beta -3\right) h _0+ 5\ 3^{1/3}\alpha ^2 \left(61 \beta ^2-130 \beta +65\right) \gamma ^{2/3} h_0^{2}\right) \notag \\
    &+\mathcal{O}(\gamma^{-1}).
    \label{Solution-k}
\end{align}
As a check to our method, we recover the homogeneous case by taking $\alpha=0$ or $\beta=1$ to obtain
\begin{align}
 \stackrel{\circ}{\lambda}_{\textrm{cr}}&=1-\frac{3^{2/3}}{4 \gamma ^{2/3}}+\frac{33\ 3^{1/3}}{160\gamma ^{4/3}}-\frac{3^{2/3}}{8\gamma ^{5/3}}+\mathcal{O}(\gamma^{-2}),
    \label{Solution-lambda-homogeneous}\\
     \stackrel{\circ}{k}_{\textrm{cr}}&=\frac{3^{1/3}}{\pi \gamma^{1/3} h _{0}} -\frac{27}{20\pi\gamma  h _0}+\mathcal{O}(\gamma^{-1}).
    \label{Solution-mode-homogeneous}
\end{align}
where an upper circle stands for the solution corresponding to a homogeneous substrate. Using \eqref{Wavenumber} the estimates \eqref{Solution-mode-homogeneous} gives 
\begin{align}
    (nh _{0})_{\textrm{cr}}=\frac{3^{1/3}}{\gamma^{1/3}}-\frac{3}{5\gamma}+\mathcal{O}(\gamma^{-5/3})\quad \textrm{or} \quad (nh )_{\textrm{cr}}=\frac{3^{1/3}}{\gamma^{1/3}}+\frac{3}{20\gamma }+\mathcal{O}(\gamma ^{-5/3}),
    \label{Solution-k-homogeneous}
\end{align}
which is an exact match to the results found in \cite{Cai1999PRSA,Alawiye2019PTRSA,Wang2023MMS} up to $\mathcal{O}(\gamma^{-5/3})$. 

A remarkable property of the solution \eqref{Solution-lambda} and \eqref{Solution-k}, is that all parameters relevant to gradients (given by $\alpha$ and $\beta$), only appear in higher order terms. In other words, it is  the modulus ratio between the film and the substrate surface that dominates the critical bifurcation condition, and we find that the modulus gradient  has a minor influence on $\lambda_{\textrm{cr}}$ and $k_{\textrm{cr}}$.  To further unravel how the modulus gradient affects the critical state, we define
\begin{align}
\delta_1=&\stackrel{\circ}{\lambda} _{\textrm{cr}}-\lambda_{\textrm{cr}}=\frac{3^{1/3} \alpha  \left(\beta -1\right)h _{0}}{4 \gamma^{1/3} }+\mathcal{O}(\gamma^{-4/3}),\label{difference-stretch}\\
\delta_2=&\stackrel{\circ}{k} _{\textrm{cr}}-k_{\textrm{cr}}=-\frac{\alpha  \left(\beta -1\right)}{\pi}+\mathcal{O}(\gamma^{-1/3}).\label{difference-mode}
\end{align}
The sign of $\delta_1$ implies whether a bilayer with graded substrate is more stable or more unstable than the corresponding homogeneous substrate. Thus, we summarize the results of the signs of $\delta_1$ and $\delta_2$ as follows
\begin{equation}
\delta_1\quad\textrm{or}~-\delta_2~
\left\{\begin{aligned}
&>0,\quad\textrm{if}\quad \alpha<0,~\beta<1\quad \textrm{or}\quad \alpha>0,~\beta>1,\\ 
&<0,\quad\textrm{if}\quad  \alpha>0,~\beta<1.
\end{aligned}\right.
\end{equation}
Note that the prerequisite that \eqref{difference-stretch} and \eqref{difference-mode} hold is $\beta\sim\mathcal{O}(1)$. Therefore we may allow $\beta>1$ but a large $\beta$ is not granted. It can be seen that either $ \alpha<0,~\beta<1$ or $\alpha>0,~\beta>1$ indicates a positive $\delta_1$ but a negative $\delta_2$, suggesting that an increasing modulus distribution will delay surface wrinkling and lead to a larger wavenumber compared to its homogeneous counterpart. Contrariwise,  the case $\alpha>0,~\beta<1$, with an exponentially decayed modulus, always gives rise to a more unstable structure and a lower wavenumber. These conclusions further confirm previous results \cite{Cao2012IJSS}.

 \begin{figure}[!bhtp]
    \centering
    \subfigure[Dependence of $\lambda_{\textrm{cr}}$ on $\gamma$.]
            {\includegraphics[scale=0.92]{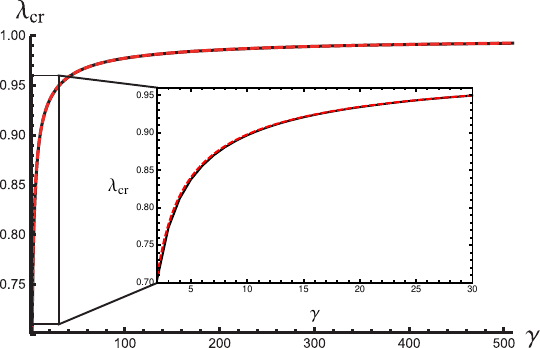}
    \label{Lambda-gamma}}
   \subfigure[Dependence of $k_{\textrm{cr}}$ on $\gamma$.]
            {\includegraphics[scale=0.92]{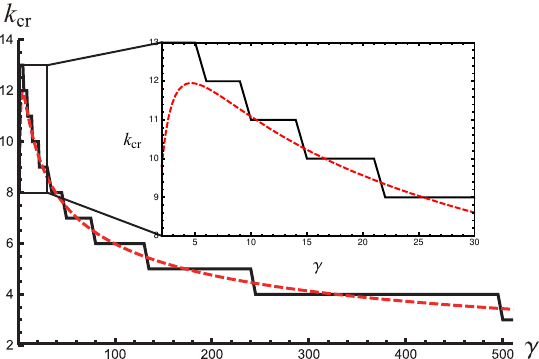}
    \label{Mode-gamma}}
    \caption{Comparisons between the asymptotic (dashed line) and exact (solid line) solutions of $\lambda_{\textrm{cr}}$ and $ k_{\textrm{cr}}$ for $h _{0}$=0.015, $\alpha=1$, $\beta=0.01$ when $\gamma\sim\mathcal{O}(h_{0}^{-\frac{3}{2}})$.}
    \label{fig6}
\end{figure}

We now check the validity our asymptotic solutions given by \eqref{Solution-lambda} and \eqref{Solution-k}. Figure \ref{fig6} show how $\lambda_{\textrm{cr}}$ and $k_{\textrm{cr}}$ depend on $\gamma$. We see that the error between asymptotic and exact results is negligible for large $\gamma$. Further, we observe that the asymptotic solution remains valid even for small $\gamma$ and at $\gamma\approx 2$ the largest error is less than $0.6\%$. We also see in Figure \ref{Mode-gamma} that  \eqref{Solution-k} looses its validity $\gamma\lessapprox5$. These solutions also 
demonstrate that a larger mismatch of the stiffness between the film and substrate always tends to destabilize the structure with a higher $\lambda_\textrm{cr}$ but a lower $k_\textrm{cr}$ which is consistent with previous results.

\begin{figure}[!htpb]
    \centering
    \subfigure[Dependence of $\lambda_{\textrm{cr}}$ on $h _0$.]
    {\includegraphics[scale=0.92]{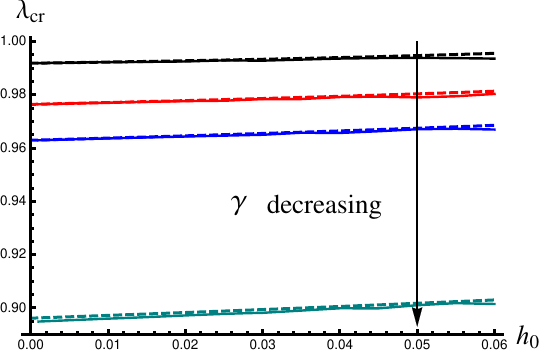}\label{Lambda-h0-fixedgamma}}
    \subfigure[Dependence of $k_{\textrm{cr}}$ on $h _0$.]
    {\includegraphics[scale=0.92]{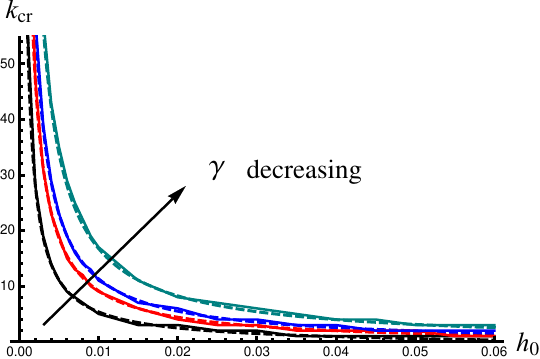} \label{k-h0-fixedgamma}}
    \caption{Comparisons between the asymptotic (dashed line) and exact (solid line) solutions of $\lambda_{\textrm{cr}}$ and $k_{\textrm{cr}}$ for $\alpha=1$, $\beta=0.01$ and $\gamma=500,~100,~50,~10$ when $\gamma\sim\mathcal{O}(h_{0}^{-\frac{3}{2}})$.}
    \label{Lambda-mode-h0}
\end{figure}

Next we investigate the effect of $h _{0}$ in Figure \ref{Lambda-mode-h0}. Again the  agreement between exact and asymptotic solutions is excellent, and we observe that unless the substrate is homogeneous \citep{Liu2014IJES}, the critical stretch is affected by $h _{0}$. Indeed,  $\lambda_{\textrm{cr}}$ in \eqref{Solution-lambda-homogeneous} is depends  on $h _{0}$ to second and higher order terms for a graded substrate but with factors  $\alpha$ and $\beta-1$ that  vanish identically for homogeneous substrates. The effect of $h _{0}$ is weak as seen in Figure \ref{Lambda-h0-fixedgamma}. This results seems unexpected at first sight as the bilayer is more unstable with a thicker film. However, since $k_{\textrm{cr}}$ is proportional to $1/h _{0}$, a thicker film develops   less wrinkles (see Figure \ref{k-h0-fixedgamma}). We emphasize that the order relation $\gamma\sim\mathcal{O}(h_{0}^{-\frac{3}{2}})$ implies that the film is much stiffer that the substrate for extremely thin films.


\subsubsection{Moderately stiff film on soft graded substrate: $\gamma\sim\mathcal{O}(h_{0}^{-1})$}\label{case2}

From the leading-order balance analysis of \eqref{Asymptotic-expansion-case1} we discover an implied relation between $\gamma$ and $kh_0 $, i.e. $kh_0 \sim\mathcal{O}(\gamma^{-\frac{1}{3}})$. We now assume that the film is not as stiff and have, instead,  $\gamma=\mathcal{O}(h_{0}^{-1})$. In this case, we have $k=\mathcal{O}(\gamma ^{\frac{2}{3}})$. and with $h _{0}=a_{0}\gamma^{-1}$,  the asymptotic solutions for $\lambda_{\textrm{cr}}$ and $k_{\textrm{cr}}$ are now 
\begin{equation}
\begin{aligned}
    &\lambda_{\textrm{cr}} =1-\varepsilon=1-m_{0}\gamma ^{-\frac{2}{3}} -m_{1}\gamma^{-1} -m_{2}\gamma ^{-\frac{4}{3}} -m_{3}\gamma ^{-\frac{5}{3}}+\dots, \\
    &k_{\textrm{cr}}=l_{0} \gamma ^{2/3}+ l_{1}\gamma ^{1/3}+l_{2} +l_{3}\gamma ^{-\frac{1}{3}} \dots,
    \label{Asymptotic-form-2}
\end{aligned}
\end{equation}
where $m_{i}$ and $l_{i}$ $(i=0, 1, 2, \dots)$ are constants depending on $h _{0}$, $\alpha$ and $\beta$ and can be solved from equations \eqref{f} and \eqref{Dfdk}. 

Following the same procedure as in the previous case, we obtain the asymptotic solutions:
\begin{align}
    &\lambda _{\textrm{cr}}=1-\frac{3^{2/3}}{4 \gamma ^{2/3}}+\frac{3^{1/3}}{160\gamma ^{4/3}} \left(33-40 \alpha  \left(\beta -1\right) \gamma  h _0 \right)-\frac{3^{2/3}}{24\gamma ^{5/3}} \left(3-2 \alpha  \left(\beta -1\right) \gamma  h _0\right)+\mathcal{O}(\gamma^{-2}), \label{Solution-lambda-1}\\
    &k_{\textrm{cr}}=\frac{3^{1/3}}{\pi \gamma^{1/3} h _{0}}+\frac{\alpha  \left(\beta -1\right)}{\pi} -\frac{27}{20 \pi \gamma  h _0}-\frac{\alpha  \left(\beta -1\right)}{2\ 3^{2/3} \pi \gamma^{1/3} }+\mathcal{O}(\gamma^{-2/3}).
    \label{Solution-k-1}
\end{align}
These solutions reduce to \eqref{Solution-lambda-homogeneous} and \eqref{Solution-mode-homogeneous}, when either $\alpha=0$ or $\beta=1$. Further, the leading order term of \eqref{Solution-lambda-1} and the first two terms of \eqref{Solution-k-1} are identical to those in \eqref{Solution-lambda} and \eqref{Solution-k}. Nevertheless, an interesting difference is that the lowest order where the information of modulus gradient occurs is $\mathcal{O}(\gamma^{-\frac{4}{3}})$ in \eqref{Solution-lambda-1} while the corresponding order is $\mathcal{O}(\gamma^{-1})$ in \eqref{Solution-lambda}. Hence, the influence of modulus gradient on the critical stretch $\lambda_{\textrm{cr}}$ becomes negligible as the modulus mismatch between the film and substrate drops. For the critical wavenumber, the first two terms of \eqref{Solution-k-1} are consistent with those of \eqref{Solution-k}, indicating that the buckled morphology is nearly impervious as order between $\gamma$ and $h_0 $ varies from $\gamma\sim\mathcal{O}(h_0^{-3/2})$ to $\gamma\sim\mathcal{O}(h_0^{-1})$.


\subsubsection{Slightly stiffer film on soft graded substrate: $\gamma\sim\mathcal{O}(h_{0}^{-\frac{1}{2}})$}\label{case3}
Finally, we consider the case $\gamma\sim\mathcal{O}(h_{0}^{-\frac{1}{2}})$ which implies $k\sim\mathcal{O}(\gamma^{\frac{5}{3}})$, and we obtain
\begin{align}
    &\lambda _{\textrm{cr}}=1-\frac{3^{2/3}}{4 \gamma ^{2/3}}+\frac{33\ 3^{1/3}}{160\gamma ^{4/3}}-\frac{3^{2/3}}{8\gamma ^{5/3}}+\mathcal{O}(\gamma^{-2}),\label{Asymptotic-lambda-2}\\
    &k_{\textrm{cr}}=\frac{3^{1/3}}{\pi \gamma^{1/3}h _{0}}-\frac{27}{20 \pi \gamma h _{0}}+\mathcal{O}(\gamma^{1/3}).
    \label{Asymptotic-k-2}
\end{align}
These results are independent of the modulus gradient parameters $\alpha$, $\beta$ and the thickness $h _{0}$. Specifically, equations \eqref{Asymptotic-lambda-2} and \eqref{Asymptotic-k-2} coincide exactly with \eqref{Solution-lambda-homogeneous} and $\eqref{Solution-mode-homogeneous}$, respectively, up to the truncated order. This implies that for a stiff film, the graded substrate  makes little difference compared to its homogeneous counterpart. 

Combining the results for $\gamma\sim\mathcal{O}(h_{0}^{-3/2})$ and $\gamma\sim\mathcal{O}(h_{0}^{-1})$, we conclude that for a film/substrate bilayer, where the film thickness is fixed and the shear modulus of the substrate either decays exponentially from the surface or is comparable to the modulus at infinity, the influence of the modulus gradient tends to shift to higher orders as the modulus ratio $\gamma$ decreases, playing a more and more marginal role in regulating surface wrinkling as well as in shaping the final surface pattern.


\subsection{Restricted growth}
For a stiff layer coated to a soft substrate, it is known  that the critical growth factor $g_{\textrm{cr}}$ leading to wrinkling is close to unity \citep{Alawiye2019PTRSA}. Therefore,  $\varepsilon=g_{\textrm{cr}}-1$ is small. Following the fundamental assumption in Section \ref{Uniaxial-compression1}, we assume that $\alpha$ is of $\mathcal{O}(1)$ and $\varepsilon, 1/k, 1/\gamma, h _{0}$ are small. 

\begin{figure}[!htp]
    \centering
    \subfigure[$\gamma=$ 500, 200, 60, 25, 10.]
    {
            \includegraphics[scale=0.92]{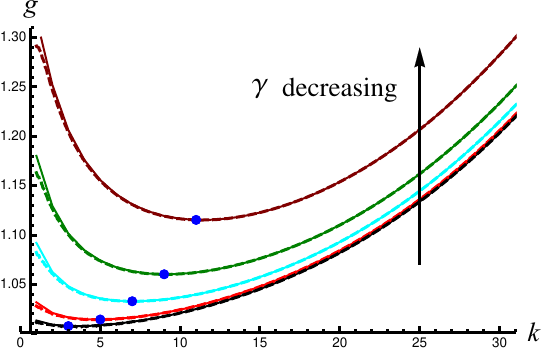}
    \label{Growth-mode-fixedh0}
    }
    \subfigure[$h _{0}=$ 0.06, 0.03, 0.015, 0.01, 0.008, 0.006, 0.005.]
    {
            \includegraphics[scale=0.92]{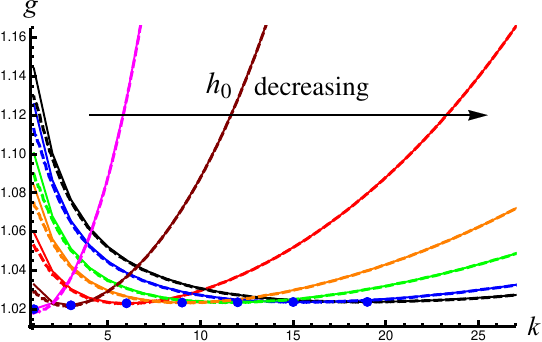}

    \label{Growth-mode-fixedgamma}
    }
    \caption{Comparisons of the bifurcation curves between the approximate (dashed line) and exact (solid line) bifurcation conditions when $\alpha=1$ and $\beta=0.01$. The left subfigure displays the dependence of $g$ on $k$ by specifying $h_0 =0.015$ while the right one exhibits the counterpart for fixed $\gamma=100$. The blue dots correspond to the critical growth factor $g_{\textrm{cr}}$ associated with the critical wavenumber $k_{\textrm{cr}}$.}
    \label{Growth-mode}
\end{figure}

To evaluate the accuracy of the approximate bifurcation condition, we compare in Figure \ref{Growth-mode} the  the bifurcation curves  the approximate (dashed line) and exact (solid line) bifurcation conditions. The dashed curves are an excellent fit for the solids curves, which indicates the validity of the explicit bifurcation as our new starting point to pursue an asymptotic solution.  In Figure \ref{Growth-mode} we see that  the bifurcation curves have a $U-$shape. The  critical growth factor $g_{\textrm{cr}}$ and the critical wavenumber $k_{\textrm{cr}}$ correspond to the vertical and horizontal coordinates of the local minimum for each curve. 


\subsubsection{Very stiff film on soft graded substrate: $\gamma\sim\mathcal{O}(h^{-\frac{3}{2}}_{0})$}
We first expand the bifurcation condition $\bar{\Xi}=0$ in $\varepsilon$ and $h _{0}$. Referring to the order analysis in Section \ref{case1} and by using the leading-order balance and principle of least degeneracy \citep{Liu2014IJES}, we arrive at the same order relations as before, namely, $\varepsilon\sim\mathcal{O}(\gamma^{-\frac{2}{3}})$, $h _{0}\sim\mathcal{O}(\gamma^{-\frac{2}{3}})$ and $k\sim\mathcal{O}(\gamma^{\frac{1}{3}})$. Therefore, we look for an asymptotic solution of the form
\begin{equation}
\begin{aligned}
    &g_{\textrm{cr}} =1+\varepsilon=1+m_{0}\gamma ^{-\frac{2}{3}} +m_{1}\gamma^{-1} +m_{2}\gamma ^{-\frac{4}{3}} +m_{3}\gamma ^{-\frac{5}{3}}+\dots,\\
    &k_{\textrm{cr}}=l_{0} \gamma ^{1/3}+ l_{1}+l_{2}\gamma ^{-\frac{1}{3}} +l_{3}\gamma ^{-\frac{2}{3}} \dots,
    \label{Asymptotic-form-growth}
\end{aligned}
\end{equation}
where $m_i$ and $l_i$ $(i=0,1,2,\cdots)$ are constants to be determined.

Following the same procedure given in Section \ref{case1}, we  derive  the following formulas
\begin{align}
    g_{\textrm{cr}}=&~1+\frac{3^{2/3}}{4 \gamma ^{2/3}} +\frac{3^{1/3} \alpha  (\beta -1) h _0}{4 \gamma^{1/3}} -\frac{1}{480}\left(\frac{9\ 3^{1/3}}{\gamma ^{4/3}} +\frac{40\ 3^{2/3} \alpha  (\beta -1) h _0}{\gamma ^{2/3}} +100 \alpha ^2 (\beta -1)^2 h^{2}_0\right) \notag\\
    &+\frac{1}{720} \bigg( \frac{90\ 3^{2/3}}{\gamma ^{5/3}} +\frac{81 \alpha  \left(\beta -1\right) h _0}{\gamma } +\frac{30\ 3^{1/3} \alpha ^2 \left(\beta ^2-3 \beta +2\right) h_0^{2}}{\gamma^{1/3}} \notag\\
    &+10\ 3^{2/3} \alpha ^3 \left(10 \beta ^3-32 \beta ^2+33 \beta -11\right) \gamma^{1/3} h_0^{3}\bigg)+\mathcal{O}(\gamma^{-2}), \label{Solution-g-growth}\\
    k_{\textrm{cr}}=&~\frac{3^{1/3}}{\pi \gamma^{1/3} h _{0}} +\frac{\alpha  \left(\beta -1\right)}{\pi} -\frac{1}{60 \pi}\left( \frac{81}{\gamma  h _0} +\frac{10\ 3^{1/3} \alpha  \left(\beta -1\right)}{\gamma ^{1/3}} +35\ 3^{2/3} \alpha ^2 \left(\beta -1\right)^2 \gamma ^{1/3} h _0  \right)    \notag\\
    &-\frac{\alpha  \left(\beta -1\right)}{180 \pi}\left( \frac{42\ 3^{2/3}}{\gamma ^{2/3}} -30 \alpha  \left(2 \beta -3\right) h _0 -5\ 3^{1/3}\alpha ^2 \left(61 \beta ^2-130 \beta +65\right) \gamma ^{2/3} h_0^{2} \right)  \notag \\
    &+\mathcal{O}(\gamma^{-1}).
    \label{Solution-k-growth}
\end{align}
For a homogeneous substrate, equations \eqref{Solution-g-growth} and \eqref{Solution-k-growth} simplify to 
\begin{align}
   &\stackrel{\circ}{g}_{\textrm{cr}}=1+\frac{3^{2/3}}{4 \gamma ^{2/3}}  -\frac{3\ 3^{1/3}}{160 \gamma ^{4/3}} + \frac{ 3^{2/3}}{8 \gamma ^{5/3}} +\mathcal{O}(\gamma^{-2}), \label{Solution-g-growth-homogeneous}\\
    &\stackrel{\circ}{k}_{\textrm{cr}}=\frac{3^{1/3}}{\pi \gamma^{1/3} h _{0}} -\frac{27}{20 \pi \gamma  h _0} +\mathcal{O}(\gamma^{-1}) \quad \textrm{or} \quad 
    (nh _{0})_{\textrm{cr}}=\frac{3^{1/3}}{\gamma^{1/3}} -\frac{27}{20 \gamma} +\mathcal{O}(\gamma^{-5/3}).
    \label{Solution-k-growth-homogeneous}
\end{align}
We emphasize that the leading order terms of \eqref{Solution-g-growth-homogeneous} and $\eqref{Solution-k-growth-homogeneous}_2$ are the same as those in \cite{Alawiye2019PTRSA}, where there is no growth in the substrate and only  film growth is considered. This reinforces our conclusion in \cite{Liu2021PRSA} that growth gradient has a little effect on the critical buckling load and the associated buckling pattern. Further, it can be seen that the thickness $h _{0}$ and the parameters related to modulus gradient $\alpha$, $\beta$ are included at higher order terms. So the dominant factor that affects the critical buckling state is still $\gamma$. The same conclusion has already been observed in the compression case.

\begin{figure}[htp]
    \centering
    \subfigure[Dependence of $g_{\textrm{cr}}$ on $\gamma$.]
    { 
            \includegraphics[scale=0.92]{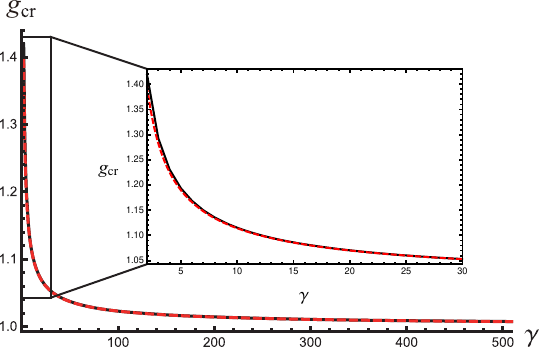}
    \label{Growth-gamma}
    }
    \subfigure[Dependence of $k_{\textrm{cr}}$ on $\gamma$.]
    {
            \includegraphics[scale=0.92]{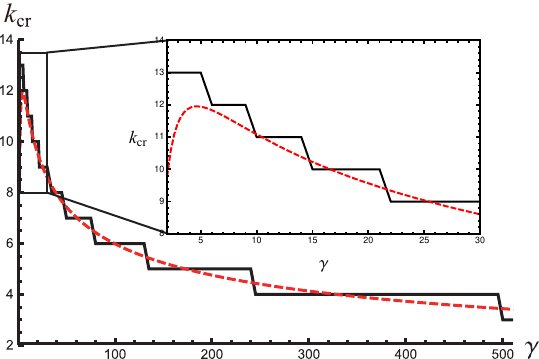}
    \label{Mode-gamma-growth}
    }
    \caption{Comparisons between the asymptotic (dashed line) and exact (solid line) solutions of  $g_{\textrm{cr}}$ and $k_{\textrm{cr}}$ for $\alpha=1$, $\beta=0.01$, and $h _{0}=0.015$ when $\gamma\sim\mathcal{O}(h^{-\frac{3}{2}}_{0})$.}
    \label{Comparison-growth}
\end{figure}

Figure \ref{Comparison-growth} illustrates the critical growth factor $g_{\textrm{cr}}$ and the critical wavenumber $k_{\textrm{cr}}$ as functions of the modulus ratio $\gamma$ when $\alpha=1$, $\beta=0.01$, and $h _0=0.015$. The dashed lines correspond to the asymptotic solution while the solid lines represent the exact solution. We note the surprisingly good agreement between the exact and asymptotic solutions, even when $\gamma\approx 5$. It implies that the general order relation $\varepsilon\sim\mathcal{O}(\gamma^{-\frac{3}{2}})$ captures well yje surface instability in hyperelastic bilayers.

\begin{figure}[htp]
    \centering
    \subfigure[Dependence of $g_{\textrm{cr}}$ on $h _{0}$.]
    {
            \includegraphics[scale=0.92]{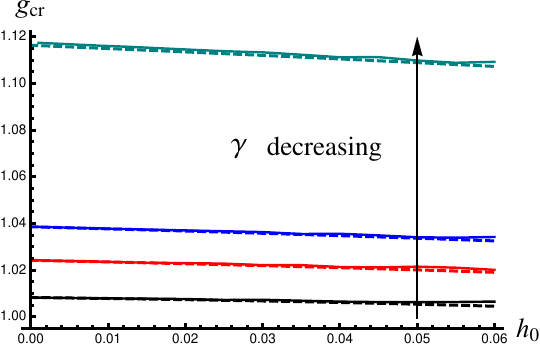}
    \label{Growth-h0}
    }
    \subfigure[Dependence of $k_{\textrm{cr}}$ on $h _{0}$.]
    {
            \includegraphics[scale=0.92]{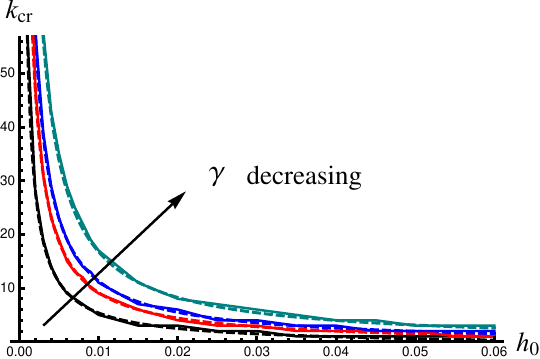}
    \label{Mode-h0-growth}
    }
    \caption{Comparisons between the asymptotic (dashed line) and exact (solid line) solutions of $g_{\textrm{cr}}$ and $k_{\textrm{cr}}$ for $\alpha=1$, $\beta=0.01$ and $\gamma=500, 100, 50, 10$ when $\gamma\sim\mathcal{O}(h_{0}^{-\frac{3}{2}})$.}
    \label{Comparison-growth-1}
\end{figure}

We plot the relations of $g_{\textrm{cr}}$, $k_{\textrm{cr}}$ and $h_0 $ in Figure \ref{Comparison-growth-1} to further validate our asymptotic solutions. We found from the second order term in \eqref{Solution-g-growth} that the film thickness $h_0 $ starts to affect the critical growth factor $g_{\textrm{cr}}$, as confirmed in Figure \ref{Growth-h0}. Note that $h_0 $ does not impact the onset of surface wrinkling for a homogeneous substrate  through  \eqref{Solution-k-growth}, it will change  the wavenumber, as shown in \ref{Mode-h0-growth}. We conclude that the  asymptotic solutions are excellent approximations to describe the onset of surface instability and the corresponding pattern.

\subsubsection{Moderately stiff film on soft graded substrate: $\gamma\sim\mathcal{O}(h^{-1}_{0})$}
When $\gamma\sim\mathcal{O}(h^{-1}_{0})$, $k\sim\mathcal{O}(\gamma^\frac{2}{3})$ we obtain the following asymptotic solutions
\begin{align}
    &g_{\textrm{cr}}=1+\frac{3^{2/3}}{4 \gamma ^{2/3}}-\frac{3^{1/3}}{160\gamma ^{4/3}} \left(3 +40 \alpha  \left(\beta -1\right) \gamma  h _0 \right)+\frac{3^{2/3}}{24\gamma ^{5/3}} \left(3-2 \alpha  \left(\beta -1\right) \gamma  h _0\right)+\mathcal{O}(\gamma^{-2}),     \label{Solution-g-growth2}\\
    &k_{\textrm{cr}}=\frac{3^{1/3}}{\pi \gamma^{1/3} h _{0}}+\frac{\alpha  \left(\beta -1\right)}{\pi} -\frac{27}{20 \pi \gamma  h _0}-\frac{\alpha  \left(\beta -1\right)}{2\ 3^{2/3} \pi \gamma^{1/3} }+\mathcal{O}(\gamma^{-2/3}).
    \label{Solution-k-growth2}
\end{align}
We note that $\alpha$ and $\beta$ first appear at $\mathcal{O}(\gamma^{-4/3})$, which is higher than the counterpart in \eqref{Solution-g-growth}. Hence, the effect of a modulus gradient  becomes weaker as the modulus ratio $\gamma$ declines. 


\subsubsection{Slightly stiff film on soft graded substrate: $\gamma\sim\mathcal{O}(h^{-\frac{1}{2}}_{0})$}
The last case, $\gamma\sim\mathcal{O}(h^{-\frac{1}{2}}_{0})$ and $k\sim\mathcal{O}(\gamma^{\frac{5}{3}})$ corresponds to a slightly stiff film coated to a soft graded substrate, for which, we have \begin{align}
    &g_{\textrm{cr}}=1+\frac{3^{2/3}}{4 \gamma ^{2/3}}-\frac{3^{4/3}}{160\gamma ^{4/3}}+\frac{3^{2/3}}{8\gamma ^{5/3}}+\mathcal{O}(\gamma^{-2}),\label{Solution-g-growth3}\\
    &k_{\textrm{cr}}=\frac{3^{1/3}}{\pi \gamma^{1/3}h _{0}}-\frac{27}{20\pi \gamma h _{0}}+\mathcal{O}(\gamma^{1/3}),
    \label{Solution-k-growth3}
\end{align}
which does not depend on $\alpha$ and $\beta$. Meanwhile, equations \eqref{Solution-g-growth3} and \eqref{Solution-k-growth3} are identical to \eqref{Solution-g-growth-homogeneous} and $\eqref{Solution-k-growth-homogeneous}_1$ up to the retained order, respectively, implying that a homogeneous substrate can indeed replace a graded one when estimating the critical buckling threshold and the associated buckling pattern. In addition, we give the performance of the asymptotic solutions \eqref{Solution-g-growth}, \eqref{Solution-g-growth2} and \eqref{Solution-g-growth3} when $\gamma$ is slightly greater than unity in Table \ref{table1}. Again, the relative errors are small even when $\gamma\approx 2$. The solution of $\gamma=\mathcal{O}(h^{-\frac{1}{2}}_{0})$ yields the best prediction.

We conclude that in the confined growth scenario when both $\alpha$ and $\beta$ are of $\mathcal{O}(1)$ and both $h_0 $ and $1/\gamma$ are small, the influence of the modulus gradient on pattern formation  in a stiff film boned to a compliant graded substrate  is negligible. For smaller $\gamma$, the modulus gradient becomes irrelevant.

\begin{table}[h]
    \centering
    \caption{Comparisons between the exact and asymptotic solutions of $g_{\textrm{cr}}$ as $\gamma$ is not large. All other parameters are specified by $h _{0}=0.015$, $\alpha=1$ and $\beta=0.01$. We denote $\Delta$ the relative error between the exact and asymptotic solutions.}
    \bigskip
    \begin{tabular}{c|c|cc|cc|cc}
    \toprule
    $\gamma$&Exact $g_{\textrm{cr}}$&$\gamma=\mathcal{O}(h^{-\frac{3}{2}}_{0})$&$\Delta$&$\gamma=\mathcal{O}(h^{-1})$&$\Delta$&$\gamma=\mathcal{O}(h^{-\frac{1}{2}}_{0})$&$\Delta$\\
    \hline
    9&1.12453&1.12322&0.11682\%&1.12344&0.09717\%&1.12542&-0.07883\%\\
    8&1.1357&1.13416&0.13520\%&1.13441&0.11375\%&1.13644&-0.06531\%\\
    7&1.14986&1.14787&0.17289\%&1.14814&0.14918\%&1.15024&-0.03307\%\\
    6&1.16846&1.16566&0.23979\%&1.16597&0.21312\%&1.16813&0.02766\%\\
    5&1.19371&1.18985&0.32358\%&1.19022&0.2929\%&1.19247&0.10434\%\\
    4&1.23101&1.22511&0.47965\%&1.22556&0.44322\%&1.22791&0.2522\%\\
    3&1.29281&1.28236&0.80875\%&1.28294&0.76343\%&1.28542&0.57199\%\\
    2&1.41873&1.39527&1.65369\%&1.39613&1.59297\%&1.39876&1.40772\%\\
    \bottomrule
    \end{tabular}\\[10pt]
    \label{table1}
\end{table}


\section{Asymptotic solution when $\beta$ is large}\label{Asymptotic-analysis-grow}

When the modulus ratio $\beta$ is large we use the WKB approximation \eqref{WKB-expansion}. Following the method given in  Section \ref{Uniaxial-compression}, we obtain 
\begin{align}
    &S_1 = -\frac{\alpha  (1-\beta ) \lambda  \textrm{exp} \left(-\alpha  \left(\lambda  y -h _0\right)\right)}  {2 \left(\beta +(1-\beta ) \textrm{exp} \left(-\alpha  \left(\lambda  y -h _0\right)\right)\right)}\label{S1}\\&
    S_{1}' =-\frac{\alpha ^2 (\beta -1) \beta  \lambda ^2 \textrm{exp} \left(-\alpha  \left(\lambda  y -h _0\right)\right)}{2 \left(\beta  \left(\textrm{exp} \left(-\alpha  \left(\lambda  y -h _0\right)\right)-1\right)+1\right){}^2}.\label{S1prime}
\end{align}
Both $S_1$ and $S_{1}'$ are  negative under the assumption $\beta \gg 1$ and $\alpha > 0$ which implies that $S_1$ attains its local maximum at $y =h _{0}/\lambda$ with  extreme value given by
\begin{align}
    S_{1\textrm{max}}=\frac{1}{2} \alpha  (\beta -1).\label{MaxS1}
\end{align}
It can be estimated from \eqref{S1} and \eqref{MaxS1} that a boundary layer  exists after which $S_1$  decreases. However, near the surface $y =h _{0}/\lambda$, it is of $\mathcal{O}(\beta)$. The first term in \eqref{Foursol} ($kS_0$) is of $\mathcal{O}(\gamma^{\frac{1}{3}})$ and the second term ($S_1$) is of $\mathcal{O}(\beta)$. Bearing in mind that a valid asymptotic expansion requires that the latter terms are strictly smaller than the previous one, we  consider a limit case $\beta=\mathcal{O}(\gamma^{\frac{1}{3}})$, which means that the first and second terms are of the same order $\mathcal{O}(\gamma^{\frac{1}{3}})$ in the boundary layer, and the first term is much larger outside the boundary layer. Hence, we write
\begin{align}
    \beta=b_0 \gamma^{\frac{1}{3}},
\end{align}
where $b_0$ is a constant of $\mathcal{O}(1)$. We seek an asymptotic solution for $\lambda_{\textrm{cr}}$ and $k_\textrm{cr}$ the same as \eqref{Asymptotic-form}. By applying the regular solution procedure, it is found that the coefficients $l_0$ and $m_0$ satisfy the following algebraic equations
\begin{align}
    &2 \pi ^6 a_0^3 l_0^6-24 \pi ^4 a_0 m_0 l_0^4+12 \pi ^3 l_0^3+18 \pi ^2 \alpha  b_0 l_0^2+3 \pi  \alpha ^2 b_0^2 l_0-3 \alpha ^3 b_0^3=0,\label{Six-order1}\\
    &16 \pi ^6 a_0^3 l_0^6-144 \pi ^4 a_0 m_0 l_0^4+60 \pi ^3 l_0^3+72 \pi ^2 \alpha  b_0 l_0^2+9 \pi  \alpha ^2 b_0^2 l_0-6 \alpha ^3 b_0^3=0\label{Six-order2},
\end{align}
where $a_0$ has been defined in \eqref{h0}.

Unfortunately, equations \eqref{Six-order1} and \eqref{Six-order2} give rise to a polynomial equation of degree six  in $l_0$, which has no general solution. To overcome this difficulty, we resort to curve fitting using a specific function and 
the regression analysis yields
\begin{align}
    &l_{0}=0.170341 \ln\left(\frac{\alpha  \beta }{\gamma^{1/3}}+0.552186\right)+\frac{0.564331}{\left( h_{0}  \gamma ^{2/3} \right)^{0.90481}}-0.00854314, \label{l0}\\
    &m_{0}=-\frac{\alpha ^3 \beta ^3}{8 \pi ^4 \gamma ^{5/3} h_{0}  l_0^4}+\frac{\alpha ^2 \beta ^2}{8 \pi ^3 \gamma ^{4/3} h_{0}  l_0^3}+\frac{3 \alpha  \beta }{4 \pi ^2 \gamma  h_{0}  l_0^2}+\frac{1}{2 \pi  \gamma ^{2/3} h_{0}  l_0}+\frac{1}{12} \pi ^2 \gamma ^{4/3} h_{0}^{2} l_0^2.\label{m0}
\end{align}
We see that the critical wavenumber is affected by the modulus ratio $\gamma$ only through a logarithm term. A  similar dependence was also observed in growing layers \citep{Ciarletta2015IJNLM,Jin2018IJES}.

\begin{figure}[!htp]
    \centering
    \subfigure[Dependence of $\lambda_{\textrm{cr}}$ on $\beta$.]
    {
            \centering
            \includegraphics[scale=0.52]{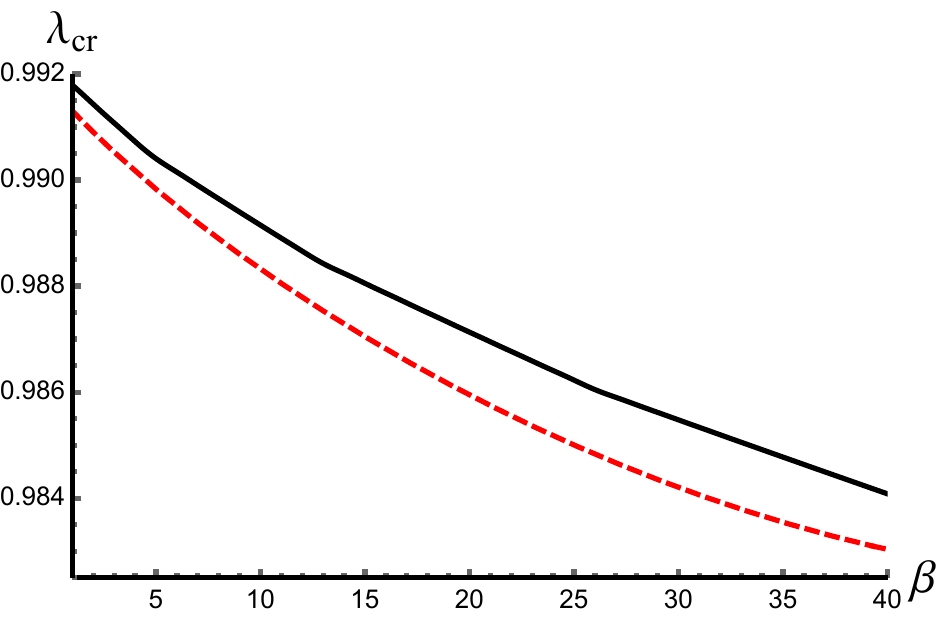}

    \label{Lambda-beta-large}
    }
    \subfigure[Dependence of $k_{\textrm{cr}}$ on $\beta$.]
    {
            \centering
            \includegraphics[scale=0.49]{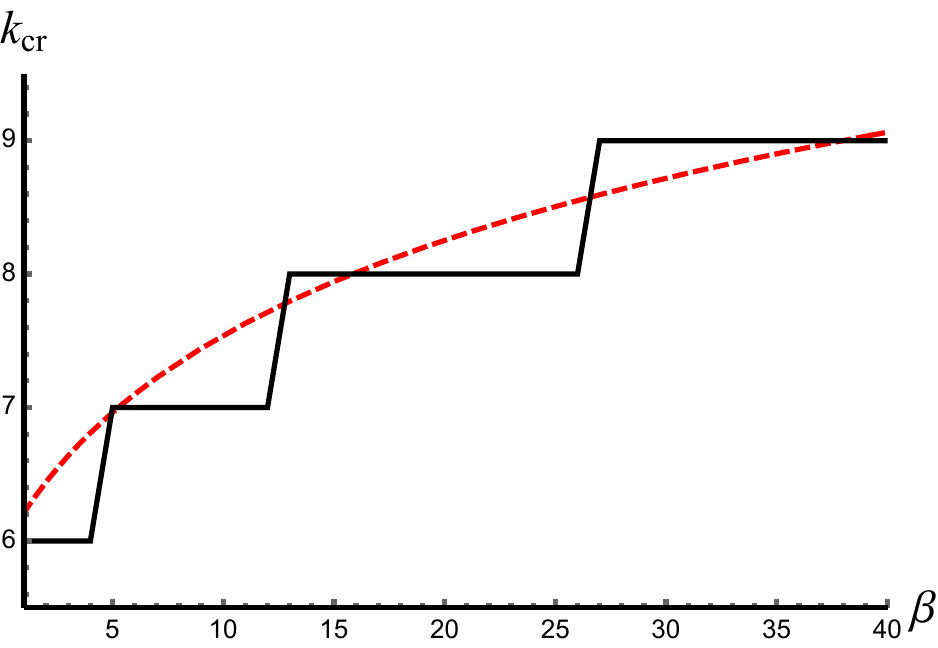}

    \label{k-beta-large}
    }
    \caption{Comparisons between the asymptotic (dashed line) and exact (solid line) solutions of $\lambda_{\textrm{cr}}$ and $ k_{\textrm{cr}}$ for $\gamma=500,h _{0}$=0.01, $\alpha=1$ when $\beta=\mathcal{O}(\gamma^{\frac{1}{3}})$.}\label{largebeta}
\end{figure}

The leading-order solutions of the critical bifurcation thresholds read $\lambda_\textrm{cr}=1-m_0/\gamma^{2/3}$ and $g_\textrm{cr}=1+m_0/\gamma^{2/3}$ while the critical wavenumber is $k_\textrm{cr}=l_0\gamma^{-1/3}$. We then illustrate the critical stretch and the critical wavenumber as functions of $\beta$ in Figure \ref{largebeta} when $\gamma=500,h _{0}$=0.01, $\alpha=1$. For comparison, both the leading-order asymptotic solution (dashed line) and the exact one (solid line) are plotted. It is emphasized that an obvious gap occurs as the vertical axial ranges from 0.983 to 0.992. In fact, the maximum error for the critical strain $1-\lambda_\textrm{cr}$ is around $5\%$ in Figure \ref{Lambda-beta-large}. On the other hand, we find that the leading-order coefficients are only valid when $\beta$ is less than 40, which is the termination of $\beta$ in Figure \ref{largebeta}. Furthermore, the critical stretch $\lambda_\textrm{cr}$ is a decreasing function of $\beta$, indicating that a greater $\beta$ tends to delay the instability. From Figure \ref{k-beta-large} we see that the critical wavenumber is an increasing function of $\beta$. So a larger $\beta$ will produce more wrinkles, which is consistent with the results in \cite{Cao2012IJSS}.

\section{Discussions and conclusions}\label{Conclusions}
We  investigated the surface wrinkling of a stiff film resting on a soft graded substrate engendered by in-plane compression or restricted growth. A theoretical model was used to identify the onset of surface wrinkling within the framework of nonlinear elasticity and by use of the incremental theory without specifying a material model and a modulus function of substrate. An exact bifurcation condition was presented by means of the Stroh formulation and the impedance matrix method \citep{Liu2022MMS,Wang2023MMS}. In the short wavelength limit, the WKB technique was applied to deal with the eigenvalue problem of ODEs with variable coefficients, and as a result an explicit bifurcation condition was obtained. By considering an exponential modulus function, the effectiveness of the approximate bifurcation condition was validated based on the exact one, and the valid domains where the WKB expansion is effective were identified for various parameters. 


We determined the order relations 
and showed that the modulus ratio $\gamma$ is extremely significant in regulating the onset of surface wrinkling in bilayer systems as found before \citep{Liu2021PRSA,Liu2022MMS} and that the critical relation $\varepsilon\sim\mathcal{O}(\gamma ^{-\frac{2}{3}})$ is consistent with most existing studies on pattern formation in film/substrate structures \citep{Cai1999PRSA,Liu2014IJES,Cai2019IJNLM,Jin2018IJES,Jia2018PRE,Wang2023MMS}. However, if surface wrinkling is replaced by a Euler-type buckling with mode number 1 or 2, the aspect ratio of the whole structure becomes dominant \citep{Liu2023AMS}. 

When $\beta$ is not large, we found that the asymptotic solutions for the critical loads and the critical wavenumber give an excellent approximation and that these solutions do not strongly depend on the modulus gradient as it only appears in higher-order terms. However, as $gamma$ decreases, and unlike an homogeneous bilayer,  the film thickness $h_0 $ starts to alter the critical load and unexpectedly a thicker film tends to destabilize the structure. 
Furthermore,  surface wrinkling is more likely if the shear modulus of the substrate decays away from the surface. 

To compare the axial compression case to the  confined growth case, we define the critical thickness $h_\textrm{cr}$ at the bifurcation point:
\begin{align}
&h _{\textrm{cr}}=h _0+\frac{3^{2/3} h _0}{4 \gamma ^{2/3}}+\frac{3^{1/3} \alpha  (\beta -1) h_0^{2}}{4 \gamma^{1/3}}+\mathcal{O}(\gamma^{-4/3}),\quad\textrm{uniaxial~compression},
\\
&h _{\textrm{cr}}=h _0+\frac{3^{2/3} h _0}{2 \gamma ^{2/3}}+\frac{3^{1/3} \alpha  (\beta -1) h_0^{2}}{2 \gamma^{1/3}}+\mathcal{O}(\gamma^{-4/3}),\quad\textrm{growth}.
\end{align}
The small difference at $\mathcal{O}(\gamma^{-2/3})$ is due to the fact that  the structure is allowed to grow in both the horizontal and the vertical directions. If we only consider  horizontal growth, the corresponding formulas would be identical up to  $\mathcal{O}(\gamma^{-4/3})$.

In the case of large $\beta$ and assuming $\beta\sim\mathcal{O}(\gamma^{1/3})$, we found the leading-order solution from a regression analysis. In this case, the parameters related to modulus gradient emerge in the leading-order terms, showing that the modulus gradient can not be neglected when the shear modulus of the substrate increases from its surface.

\begin{figure}[!htp]
    \centering
    \subfigure[Dependence of $\lambda_{\textrm{cr}}$ on $\alpha$.]
    {
            \centering
            \includegraphics[scale=0.92]{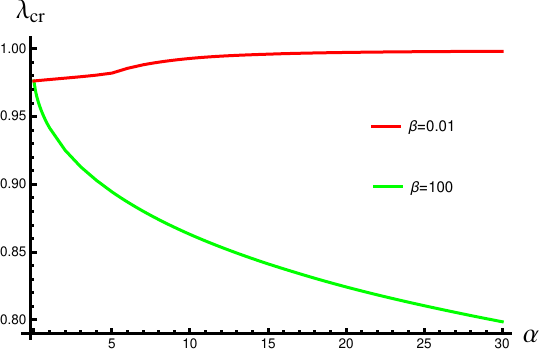}
    \label{Lambda-alpha}
    }
    \subfigure[Dependence of $k_{\textrm{cr}}$ on $\alpha$.]
    {
            \centering
            \includegraphics[scale=0.92]{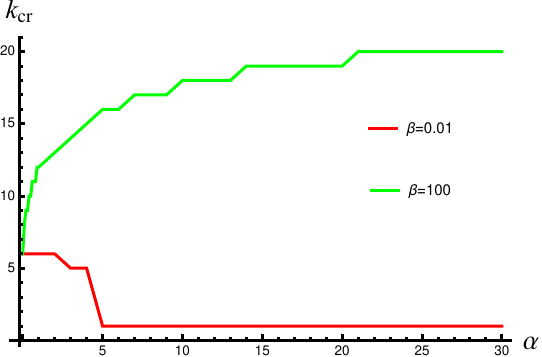}
    \label{Mode-alpha}
    }
    \caption{Influence of the grading parameter $\alpha$ on the critical stretch and the critical wavenumber with $\gamma=100$ and $h_0 =0.015$. We are concerned with the behavior as $\alpha\geqslant0$.}\label{alpha-dependence}
\end{figure}

To study the role of grading parameter $\alpha$ that controls how fast the shear modulus of the half-space decays or grows from its surface, we show in Figure~\ref{Lambda-alpha} the relation between $\lambda_\textrm{cr}$ and $\alpha$ and in Figure \ref{Mode-alpha} the relation between $k_\textrm{cr}$ and $\alpha$.
We consider two scenarios, namely, exponentially decaying modulus ($\beta=0.01$) and the reverse one ($\beta=100$).  When $\beta=0.01$, we observe from \ref{alpha-dependence} that $\lambda_\textrm{cr}$ gradually ascends but $k_\textrm{cr}$ experiences a sudden drop from $5$ to $1$ as $\alpha\approx 5$, and the $\lambda_\textrm{cr}$ curve is non-smooth as well. With increasing $\alpha$, the shear modulus of substrate will decay faster and the system will behave like triple layer structure. If a compliant substrate is coated with two stiff layers a mode transition has been observed  as the intermediate layer becomes softening  \cite{Zhou2022MMS}. This may explain the jump behavior at $\alpha\approx 5$. Indeed, a greater $\alpha$ is equivalent to reducing either the thickness or the modulus of the middle layer. In doing so, the tendency of $k_\textrm{cr}$-$\alpha$ curve in Figure \ref{Mode-alpha} (for $\beta=0.01$) is consistent with the conclusion in \cite{Cheng2014IJSS}. Moreover, the grading parameter only generates a minor correction to $\lambda_\textrm{cr}$ but creates a large change in the critical mode. When $\alpha>5$, the critical wavenumber remains one which indicates that $\alpha$ does not affect the pattern. 

When $\beta=100$, with an exponentially growing modulus in the substrate, the results are  different. Then both $\lambda_\textrm{cr}$ and $k_\textrm{cr}$ are highly dependent on $\alpha$. In particular, the critical stretch $\lambda_\textrm{cr}$ is a monotonically decreasing function of $\alpha$ while the critical wavenumber $k_\textrm{cr}$ is a monotonically increasing function of $\alpha$. As $\alpha$ goes to infinity $\lambda_\textrm{cr}$ will approach a limit value similar to the one obtained for instability of a soft wire embedded into a half-space. Indeed, for $\alpha=5000$, we find $\lambda_\textrm{cr}=0.5724$, similar to the Biot value 0.543 at which surface instability in a compressed half-space takes place \cite{Biot1963}. 


Finally, we summarize the effect of the modulus gradient. If the grading parameter $\alpha\sim\mathcal{O}(1)$, our analytical solutions show a relatively small role of the grading for an exponentially decayed modulus in the substrate as the modulus gradient only affects  high-order terms. If, however, the modulus increases away from the interface, then the gradient plays a leading role through a first-order logarithm dependence. For $\alpha$  large, corresponding to a change of modulus  in a very thin region, the grading effect has little influence on the critical state but has a very pronounced effect in shaping the pattern for an exponentially decaying modulus. Conversely, for a growing modulus, the modulus gradient has a strong effect  on both the critical stretch and the associated pattern. As $\alpha\rightarrow\infty$ the bilayer degenerates to a half-space with a soft line inclusion. Then, the classical Biot instability is recovered but with a finite wavenumber. The results presented here give a systematic way to consider wrinkling instabilities in bilayered grading material and a natural complement to computational studies.


\section*{Acknowledgment}\label{Acknowledgment}
Y.L. acknowledges the financial support from the National Natural Science Foundation of China (Project Nos 12072227 and 12021002). Y.L. and A.G. acknowledge the Guarantee Funding for Marie Sk\l odowska-Curie Actions (MSCA) Postdoctoral Fellowships (PF) 2022 from the UK Research and Innovation (EPSRC Grant No. EP/Y030559/1). The authors would like to thank the Isaac Newton Institute for Mathematical Sciences for support and hospitality during the programme Uncertainty Quantification and Stochastic Modelling of Materials when work on this paper was undertaken. This work was supported by EPSRC Grant Number EP/R014604/1. A.G. acknowledges support from the Engineering and Physical Sciences Research Council of Great Britain under Research Grant No. EP/R020205/1. For the purpose of Open Access, the author has applied a CC BY public copyright license to any Author Accepted Manuscript (AAM) version arising from this submission.

\bibliographystyle{plain}

\begin{thebibliography}{120}
    \bibitem[Alawiye et al., 2019]{Alawiye2019PTRSA}
    Alawiye, H., Kuhl, E., Goriely, A., 2019.
    Revisiting the wrinkling of elastic bilayers I: linear analysis.
    Phil. Trans. R. Soc. A. 377, 20180076.
    
    \bibitem[Alawiye et al., 2020]{Alawiye2020JMPS}
    Alawiye, H., Farrell, P. E., Goriely, A., 2020.
    Revisiting the wrinkling of elastic bilayers II: Post-bifurcation analysis.
    J. Mech. Phys. Solids 143, 104053.
    
    \bibitem[Autumn et al., 2002]{Autumn2002PNAS}
    Autumn, K., Sitti, M., Liang, Y.A., Peattie, A.M., Hansen, W.R., Sponberg, S., Kenny, T.W., Fearing, R., Israelachvili, J.N., Full, R.J., 2002.
    Evidence for van der Waals adhesion in gecko setae.
    Proc. Natl. Acad. Sci. U.S.A. 99(19) 12252--12256.
    
    \bibitem[Balbi et al., 2020]{Balbi2020PRE} 
    Balbi, V., Destrade, M., Goriely, A., 2020. 
    Mechanics of human brain organoids. 
    Phys. Rev. E. 101(2), 022403.
    
    \bibitem[Balbi et al., 2015]{Balbi2015JMPS} 
    Balbi, V., Kuhl, E., Ciarletta, P., 2015. 
    Morphoelastic control of gastro-intestinal organogenesis: Theoretical predictions and numerical insights. 
    J. Mech. Phys. Solids 78, 493--510.
    
    \bibitem[Ben Amar and Goriely, 2005]{BenAmar2005JMPS} 
    Ben Amar, M., Goriely, A., 2005. 
    Growth and instability in elastic tissues.
    J. Mech. Phys. Solids 53, 2284--2319.
    
    \bibitem[Bigoni et al., 1997]{Bigoni1997IJSS}
    Bigoni, D., Ortiz, M., Needleman, A., 1997.
    Effect of interfacial compliance on bifurcation of a layer bonded to a substrate.
    Int. J. Solids Struct. 34, 4305--4326.
    
    \bibitem[Biot, 1963]{Biot1963}
     Biot, M.A., 1963. 
     Surface instability of rubber in compression. 
     Appl. Sci. Res. 12, 168--182.
     
     
    \bibitem[Budday et al., 2015]{Budday2015}
    Budday, S., Kuhl, E., Hutchinson, J.W., 2015.
    Period-doubling and period-tripling in growing bilayered systems.
    Philos. Mag. 95, 3208--3224.
    
    \bibitem[Budday et al., 2014]{Budday2014} 
    Budday, S., Steinmann,P., Kuhl, E., 2014. 
    The role of mechanics during brain development.
    J. Mech. Phys. Solids 72, 75--92.

    \bibitem[Biryukov, 1985]{Biryukov1985} 
    Biryukov, S.V., 1985. 
    Impedance method in the theory of elastic surface waves.
    Sov. Phys. Acoust. 31, 350--354.
    
    \bibitem[Biryukov et al., 1995]{Biryukov1995} 
    Birykov, S.V., Gulyaev, Y.V., Krylov, V.V., Plessky, V.P., 1995.
    Surface acoustic waves in inhomogeneous media.
    Springer Berlin, Heidelberg.
    
    \bibitem[Cai and Fu, 1999]{Cai1999PRSA}
    Cai, Z.X., Fu, Y.B., 1999.
    On the imperfection sensitivity of a coated elastic half-space.
    Proc. R. Soc. Lond. A 455, 3285--3309.
    
    \bibitem[Cai and Fu, 2000]{Cai2000IJSS}
    Cai, Z.X., Fu, Y.B., 2000.
    Exact and asymptotic stability analyses of a coated elastic half-space.
    Int. J. Solids Struct. 37, 3101--3119.
    
    \bibitem[Cai and Fu, 2019]{Cai2019IJNLM}
    Cai, Z.X., Fu, Y.B., 2019.
    Effects of pre-stretch, compressibility and material constitution on the period-doubling secondary bifurcation of a film/substrate bilayer.
    Int. J. Non-linear Mech. 115, 11--19.
    
    \bibitem[Cao et al., 2012]{Cao2012IJSS}
    Cao, Y.P., Jia, F., Zhao, Y., Feng, X.Q., Yu, S.W., 2012.
    Buckling and post-buckling of a stiff film resting on an elastic graded substrate.
    Int. J. Solids Struct. 49, 1656--1664.
    
    \bibitem[Cao et al., 2009]{Cao2009} 
    Cao, Y.P., Zheng, X.P., Li, B., Feng, X.Q., 2009.
    Determination of the elastic modulus of micro- and nanowires/tubes using a buckling-based metrology.
    Scr. Mater. 61, 1044--1047.
    

    
    \bibitem[Chan et al., 2009]{Chan2009} 
    Chan, E.P., Page, K.A., Im, S.H., Patton, D.L., Huang, R., Stafford, C.M., 2009. 
    Viscoelastic properties of confined polymer films measured via thermal wrinkling.
    Soft Matter. 5, 4638--4641.
    
    \bibitem[Chan et al., 2008]{Chan2008AM}
    Chan, E.P., Smith, E.J., Hayward, R.C., Crosby, A.J., 2008. 
    Surface wrinkles for smart adhesion.
    Adv. Mater. 20(4), 711--716.
    
    \bibitem[Chen et al., 2017]{Chen2017PRSA}
    Chen, Z., Chen, W.Q., Song, J.Z., 2017.
    Buckling of a stiff thin film on an elastic graded compliant substrate.
    Proc. R. Soc. A 473, 20170410.
    
    \bibitem[Chen et al., 2018]{Chen2018}
    Chen, Z., Zhang, X.F., Song, J.Z., 2018.
    Surface wrinkling of an elastic graded layer.
    Soft Matter. 14, 8717--8723.
    
    \bibitem[Cheng et al., 2014]{Cheng2014IJSS}
    Cheng, H.Y., Zhang, Y.H., Huang, K.C., Rogers, J.A., Huang, Y.G., 2014.
    Buckling of a stiff thin film on a pre-strained bi-layer substrate.
    Int. J. Solids Struct. 51, 3113--3118.


    \bibitem[Ciarletta and Fu, 2015]{Ciarletta2015IJNLM} 
     Ciarletta, P., Fu, Y.B., 2015. 
     A semi-analytical approach to Biot instability in a growing layer: Strain gradient correction, weakly non-linear analysis and imperfection sensitivity.
     Int. J. Non-linear Mech. 75, 38--45.
    
    \bibitem[Coman and Destrade, 2008]{Coman2008QJMAM}
    Coman, C.D., Destrade, M., 2008.
    Asymptotic results for bifurcations in pure bending of rubber blocks.
    Q. J. Mech. Appl. Math, 61(3), 395--414.
    
    \bibitem[Dai and Liu, 2014]{Dai2014EPL}
    Dai, H.-H., Liu, Y., 2014.
    Critical thickness ratio for buckled and wrinkled fruits and vegetables.
    EPL. 108(4), 44003.
    
    \bibitem[De Pascalis et al., 2023]{DePascalis2023EML} De Pascalis, R., Lisi, F., Napoli, G., 2023.
    Solid Electrolyte Interphase elastic instability in Li-ion battery anodes.
     Extreme Mech. Lett. 61, 102014.
    
    \bibitem[Diab and Kim, 2014]{Diab2014PRSA}
    Diab, M., Kim, K.S., 2014.
    Ruga-formation instabilities of a graded stiffness boundary layer in a neo-Hookean solid.
    Proc. R. Soc. A 470, 20140218.
    
    \bibitem[Dorris and Nemat-Nasser, 1980]{Dorris1980JAM}
    Dorris, J.F., Nemat-Nasser, S., 1980.
    Instability of a layer on a half space.
    J. Appl. Mech. 47(2), 304--312.
    
    \bibitem[Eskandari et al., 2016]{Eskandari2016}
    Eskandari, M., Javili, A., Kuhl,E., 2016.
    Elastosis during airway wall remodeling explains multiple co-existing instability patterns.
    J. Theor. Biol. 403, 209--218.
    
    \bibitem[Eskandari et al., 2013]{Eskandari2013}
    Eskandari, M., Pfaller, M.R., Kuhl,E., 2013.
    On the role of mechanics in chronic lung disease. 
    Materials. 6, 5639--5658.
    
    \bibitem[Fu, 1998]{Fu1998IJNLM}
    Fu, Y.B., 1998.
    Some asymptotic results concerning the buckling of a spherical shell of arbitrary thickness.
    Int. J. Non-linear Mech. 33(6), 1111--1122.
    
    \bibitem[Fu, 2005]{Fu2005JE} 
    Fu, Y.B., 2005. 
    An integral representation of the surface-impedance tensor for incompressible elastic materials.
    J. Elast. 81, 75--90.

    \bibitem[Fu and Ciarletta, 2015]{Fu2015PRSA} 
    Fu, Y.B., Ciarletta, P., 2015. 
    Buckling of a coated elastic half-space when the coating and substrate have similar material properties.
     Proc. R. Soc. A 471, 20140979.
    
    \bibitem[Fu and Mielke, 2002]{Fu2002PRSA} 
    Fu, Y.B., Mielke, A., 2002. 
    A new identity for the surface-impedance matrix and its application to the determination of surface-wave speeds.
    Proc. R. Soc. Lond. A 458, 2523--2543.

    \bibitem[Goriely, 2017]{Gorielybook}
   Goriely, A., 2017.
   The Mathematics and Mechanics of Biological Growth.
   Springer, New York.
    
    \bibitem[Haughton and Chen, 2003]{Haughton2003ZAMP}
    Haughton, D.M., Chen, Y.C., 2003.
    Asymptotic bifurcation results for the eversion of elastic shells.
    Z. Angew. Math. Phys. 54, 191--211.
    
    \bibitem[Hinch, 1991]{Hinch1991}
    Hinch, E.J., 1991. 
    Perturbation methods. Cambridge University Press.
    
    \bibitem{hobugo18}
M.~A. Holland, S~Budday, A.~Goriely, and E.~Kuhl.
\newblock Symmetry breaking in wrinkling patterns: Gyri are universally thicker
  than sulci.
\newblock {\em Phys Rev Lett}, (In Prees), 2018.



    
    \bibitem[Huang, 2005]{Huang2005JMPS}
    Huang, R., 2005.
    Kinetic wrinkling of an elastic film on a viscoelastic substrate.
    J. Mech. Phys. Solids 53, 63--89.
    
    \bibitem[Huang and Suo, 2002]{Huang2002IJSS}
    Huang, R., Suo, Z., 2002.
    Instability of a compressed elastic film on a viscous layer.
    Int. J. Solids Struct. 39, 1791--1802.
    
    \bibitem[Hutchinson, 2013]{Hutchinson2013PTRSA}
    Hutchinson, J.W., 2013.
    The role of nonlinear substrate elasticity in the wrinkling of thin films.
    Phil. Trans. R. Soc. A. 371, 20120422.
    
    \bibitem[Im and Huang, 2008]{Im2008JMPS}
    Im, S.H., Huang, R., 2008.
    Wrinkle patterns of anisotropic crystal films on viscoelastic substrates.
    J. Mech. Phys. Solids 56, 3315--3330.
    
    \bibitem{jia2018curvature}
Fei Jia, Simon~P Pearce, and Alain Goriely.
\newblock Curvature delays growth-induced wrinkling.
\newblock {\em Physical Review E}, 98(3):033003, 2018.


    
    \bibitem[Jia et al., 2014]{Jia2014IJAM}
    Jia, F., Cao, Y.P., Zhao, Y., Feng, X.Q., 2014.
    Buckling and surface wrinkling of an elastic graded cylinder with elastic modulus arbitrarily varying along radial direction.
    Int. J. Appl. Mech. 6(1), 1450003.
    
    \bibitem[Jia et al., 2018]{Jia2018PRE} Jia, F., Pearce, S., Goriely, A., 2018.
    Curvature delays growth-induced wrinkling.
    Phys. Rev. E. 98, 033003
    
    
    \bibitem[Jin et al., 2018]{Jin2018IJES}
    Jin, L.S., Liu, Y., Cai, Z.X., 2018.
    Asymptotic solutions on the circumferential wrinkling of growing tubular tissues.
    Int. J. Eng. Sci. 128, 31--43.
    
    \bibitem[Khare et al., 2009]{Khare2009JACS}
    Khare, K., Zhou, J.H., Yang, S., 2009.
    Tunable open-channel microfluidics on soft poly(dimethyl-siloxane)(PDMS) substrates with sinusoidal grooves.
    J. Am. Chem. Soc. 25(21), 12794--12799.
    
    \bibitem[K\"ucken and Newell, 2004]{Kucken2004EPL}
    K\"ucken, M., Newell, A.C., 2004.
    A model for fingerprint formation.
    EPL. 68(1), 141--146.
    
    \bibitem[Lee et al., 2008]{Lee2008JMPS}
    Lee, D., Triantafyllidis, N., Barber, J.R., Thouless, M.D., 2008.
    Surface instability of an elastic half space with material properties varying with depth.
    J. Mech. Phys. Solids 56, 858--868.
    
    \bibitem[Lee et al., 2022]{Lee2022Small}
Lee, G., Zarei, M., Wei, Q., Zhu, Y., Lee, S.G., 2022.
Surface wrinkling for flexible and stretchable sensors.
Small. 18, 2203491.
    
    \bibitem[Lee et al., 2021]{Lee2021EML}
Lee, J., Jung, S., Kim, W., 2021.
Dependence of the effective surface tension of liquid phase eutectic gallium indium on wrinkles of the surface oxide.
Extreme. Mech. Lett. 48, 101386.
    
    \bibitem[Lee et al., 2010]{Lee2010AM}
    Lee, S.G., Lee, D.Y., Lim, H.S., Lee, D.H., Lee, S., Cho, K., 2010. 
    Switchable transparency and wetting of elastomeric smart windows.
    Adv. Mater. 22(44), 5013--5017.
    
    \bibitem[Li et al., 2011a]{Li2011JB}
    Li, B., Cao, Y.P., Feng, X.Q., 2011.
    Growth and surface folding of esophageal mucosa: A biomechanical model.
    J. Biomech. 44, 182--188.
    
    \bibitem[Li et al., 2012]{Li2012SM} Li, B., Cao, Y.P., Feng, X.Q., Gao H., 2012.Mechanics of morphological instabilities and surface wrinkling in soft materials: a review.
    Soft Matter 8, 5728.
    
    \bibitem[Li et al., 2011b]{Li2011JMPS}
    Li, B., Cao, Y.P., Feng, X.Q., Gao, H., 2011.
    Surface wrinkling of mucosa induced by volumetric growth: Theory, simulation and experiment.
    J. Mech. Phys. Solids. 59, 758--774.
    
    \bibitem[Li et al., 2017]{Li2017}
    Li, Z.W., Zhai, Y., Wang, Y., Wendland, G.M., Yin, X.B., Xiao, J.L., 2017. 
    Harnessing surface wrinkling-cracking patterns for tunable optical transmittance.
    Adv. Opt. Mater. 5(19), 1700425.
    
    \bibitem[Liu et al., 2021]{Liu2021PRSA}
    Liu, R.C., Liu, Y., Cai, Z.X., 2021.
    Influence of the growth gradient on surface wrinkling and pattern transition in growing tubular tissues.
    Proc. R. Soc. A 477, 20210441.
    
    \bibitem[Liu et al., 2022]{Liu2022MMS} 
    Liu, R.C., Jin, L.S., Liu, Y., Cai, Z.X., 2022.
    An experimental study of morphological formation in bilayered tubular structures driven by swelling/growth.
    Math. Mech. Solids. 27(8), 1569--1591.
    
    \bibitem[Liu, 2018]{Liu2018IJNLM}
    Liu, Y., 2018.
    Axial and circumferential buckling of a hyperelastic tube under restricted compression.
    Int. J. Non-linear Mech. 98, 145--153.
    
    \bibitem[Liu, 2023]{Liu2023AMS}
    Liu, Y., 2023.
    Higher order solution to the Euler buckling threshold for compressible hyperelastic bilayers.
    Acta Mech. Sin. 39, 422379.
    
    \bibitem[Liu and Dai, 2014]{Liu2014IJES}
    Liu, Y., Dai, H.-H., 2014.
    Compression of a hyperelastic layer-substrate structure: Transitions between buckling and surface modes.
    Int. J. Eng. Sci. 80, 74--89.
    
    \bibitem[Liu et al., 2020]{Liu2020IJNLM}
    Liu, Y., Zhang, Z., Devillanova, G., Cai, Z., 2020.
    Surface instabilities in graded tubular tissues induced by volumetric growth.
    Int. J. Non-linear Mech. 127, 103612.
    
    \bibitem[Ogden and Sotiropoulos, 1996]{Ogden1996}
    Ogden, R.W., Sotiropoulos, D.A., 1996.
    The effect of pre-stress on guided ultrasonic waves between a surface layer and a half-space.
    Ultrasonics. 34, 491--494.

    \bibitem[Moulton and Goriely, 2011]{Moulton2011JMPS}
    Moulton D.E., Goriely, A., 2011.
    Circumferential buckling instability of a growing cylindrical tube.
    J. Mech. Phys. Solids 59, 525--537.

    \bibitem[Nguyen et al., 2020]{Nguyen2020BMM}
    Nguyen, N., Nath, N., Deseri, L., Edith Tzeng, E., Velankar, S.S., Pocivavsek, L., 2020.
    Wrinkling instabilities for biologically relevant fiber‐reinforced composite materials with a case study of Neo‐Hookean/ Ogden–Gasser–Holzapfel bilayer.
    Biomech. Model.  Mechan. 19, 2375--2395.
    
    \bibitem[Qi et al., 2018]{Qi2018ACS}
    Qi, L., Ruck, C., Spychalski, G., King, B., Wu, B.X., Zhao, Y., 2018.
    Writing wrinkles on poly(dimethylsiloxane)(PDMS) by surface oxidation with a $\textrm{CO}_{2}$ laser engraver .
    ACS Appl. Mater. Interfaces. 10, 4295--4304.
    
    \bibitem[Rodriguez et al., 1994]{Rodriguez1994JB} 
    Rodriguez, E.K., Hoger, A., McCulloc, A.D., 1994. 
    Stress-dependent finite growth in soft elastic tissues.
    J. Biomech. 27(4), 455--467.
    
    \bibitem[Sabbah et al., 2016]{Sabbah2016AS} 
    Sabbah, A., Youssef, A., Damman, P., 2016. 
    Superhydrophobic surfaces created by elastic instability of PDMS. 
    Appl. Sci. 6(5), 152.
    
    \bibitem[Sanjarani Pour, 2010]{Sanjarani2010IJAM}
    Sanjarani Pour, M., 2010.
    WKB analysis of the buckling of a neo-Hookean cylindrical shell of arbitrary thickness subject to an external pressure.
    Int. J. Appl. Mech. 2(4), 857--870.
    
    \bibitem[Sanjarani Pour and Fu, 2002]{Sanjarani2002SIAM}
    Sanjarani Pour, M., Fu, Y.B., 2002. 
    WKB method with repeated roots and its application to the buckling analysis of an everted cylindrical tube.
    SIAM J. Appl. Math. 62(6), 1856--1871.
    
    \bibitem[Sanjarani Pour et al., 2013]{Sanjarani2013IJES}
    Sanjarani Pour, M., Hatami, A., Abdolalian, N., 2013.
    Another approach of WKB method for the stability analysis of the bending of an elastic rubber block.
    Int. J. Eng. Sci. 62, 1--8.
    
    \bibitem[Shield et al., 1994]{Shield1994JAM}
    Shield, T.W., Kim, K.S., Shield, R.T., 1994.
    The buckling of an elastic layer bonded to an elastic substrate in plane strain.
    J. Appl. Mech. 61(2), 231--235. 
    
    \bibitem[Shuvalov, 2003a]{Shuvalov2003PRSA} 
    Shuvalov, A.L., 2003.
    A sextic formalism for three-dimensional elastodynamics of cylindrically anisotropic radially inhomogeneous materials.
    Proc. R. Soc. Lond. A 459, 1611--1639.
    
    \bibitem[Shuvalov, 2003b]{Shuvalov2003QJMAM} 
    Shuvalov, A.L., 2003.
    The frobenius power series solution for cylindrically anisotropic radially inhomogeneous elastic materials.
    Q. J. Mech. Appl. Math, 56(3), 327--345.
    
    \bibitem[Song et al., 2008]{Song2008IJSS}
    Song, J.T., Jiang, H.Q., Liu, Z.J., Khang, D.Y., Huang, Y., Rogers, J.A., Lu, C., Koh, C.G., 2008.
    Buckling of a stiff thin film on a compliant substrate in large deformation.
    Int. J. Solids Struct. 45, 3107--3121.
    
    \bibitem[Stafford et al., 2004]{Stafford2004NM} 
    Stafford, C.M., Harrison, C., Beers, K.L., Karim, A., Amis, E.J., Vanlandingham, M.R., Kim, H., Volksen, W., Miller, R.D., Simonyi, E.E., 2004. 
    A buckling-based metrology for measuring the elastic moduli of polymeric thin films. 
    Nat. Mater. 3, 545--550.
    
    \bibitem[Steigmann and Ogden, 1997]{Steigmann1997PRSA}
    Steigmann, D.J., Ogden, R.W., 1997.
    Plane deformations of elastic solids with intrinsic boundary elasticity.
    Proc. R. Soc. Lond. A 453, 853--877.
    
    \bibitem{stewart2016wrinkling}
P.~S. Stewart, S.~L. Waters, T.~El~Sayed, D.~Vella, and A.~Goriely.
\newblock Wrinkling, creasing, and folding in fiber-reinforced soft tissues.
\newblock {\em Extreme Mechanics Letters}, 2016.
    
    \bibitem[Stroh, 1962]{Stroh1962JMP} 
    Stroh, A.N., 1962. 
    Steady state problems in anisotropic elasticity.
    J. Math. Phys. 41, 77--103.
    
    \bibitem[Su, 2020]{Su2020IJES} 
    Su, Y.P., 2020.
    Voltage-controlled instability transitions and competitions in a finitely deformed dielectric elastomer tube.
    Int. J. Eng. Sci. 157, 103380.
    
    \bibitem[Su et al., 2019]{Su2019IJSS} 
    Su, Y.P., Wu, B., Chen, W.Q., Destrade, M., 2019.
    Finite bending and pattern evolution of the associated instability for a dielectric elastomer slab.
    Int. J. Solids Struct. 158, 191--209.
    
    \bibitem[Su et al., 2016]{Su2016IJSS} 
    Su, Y., Zhou, W.J., Chen, W.Q., L\"u, C.F., 2016.
    On buckling of a soft incompressible electroactive hollow cylinder.
    Int. J. Solids Struct. 97--98, 400--416.
    
     
    \bibitem[van den Ende et al., 2013]{van2013AM}
    van den Ende, D.A., Kamminga, J.D., Boersma, A., Andritsch, T., Steeneken, P.G., 2013. 
    Voltage-controlled surface wrinkling of elastomeric coatings.
    Adv. Mater. 27, 489--512.
    
    \bibitem[Wang et al., 2020]{Wang2020IJSS}
    Wang, C.J., Zhang, S., Nie, S., Su, Y.P., Chen, W.Q., Song, J.Z., 2020.
    Buckling of a stiff thin film on a bi-layer compliant substrate of finite thickness.
    Int. J. Solids Struct. 188--189, 133--140.
    
    \bibitem[Wang et al., 2023]{Wang2023MMS}
    Wang, G., Liu, Y., Fu, Y.B.,2023.
    A refined model for the buckling of film/substrate bilayers.
    Math. Mech. Solids. 28(1), 313--330.
    
    \bibitem[Wang et al., 2016]{Wang2016APL}
    Wang, J.W., Li, B., Cao, Y.P., Feng, X.Q., Gao, H., 2016.
    Wrinkling micropatterns regulated by a hard skin layer with a periodic stiffness distribution on a soft material.
    Appl. Phys. Lett. 108, 021903.
    
    \bibitem[Wang et al., 2021]{Wang2021PAT}
Wang, M., Mu, L., Zhang, H., Ma, S., Liang, Y., Ren, L., 2021.
Flexible strain sensor with ridge-like microstructures for wearable applications.
Polym. Adv. Technol. 33, 96--103.
    
    \bibitem[Wilder et al., 2006]{Wilder2006JACS} 
    Wilder, E.A., Guo, S., Lin-Gibson, S., Fasolka, M.J., Stafford, C.M., 2006.
    Measuring the modulus of soft polymer networks via a buckling-based metrology. 
    J. Am. Chem. Soc. 39(12), 4138--4143.
    
    \bibitem[Wolfram Research Inc., 2019]{Wolfram2019}
    Wolfram Research Inc., 2019.
    Mathematica: version 12. Champaign, IL: Wolfram Research Inc.
    
    \bibitem[Wu et al., 2013]{Wu2013IJSS}
    Wu, Z.G., Bouklas, N., Huang, R., 2013.
    Swell-induced surface instability of hydrogel layers with material properties varying in thickness direction.
    Int. J. Solids Struct. 50, 578--587.
    
    \bibitem[Wu et al., 2014]{Wu2014JAM}
    Wu, Z.G., Meng, J.X., Liu, Y.H., Li, H., Huang, R., 2014.
    A state space method for surface instability of elastic layers with material properties varying in thickness direction.
    J. Appl. Mech. 81(8), 081003.
    
    \bibitem[Yang and Chen, 2017]{Yang2017PRSA}
    Yang, S.Y., Chen, Y.C., 2017.
    Wrinkle surface instability of an inhomogeneous elastic block with graded stiffness.
    Proc. R. Soc. A 473, 20160882.
    
    \bibitem[Yang et al., 2010]{Yang2010AFM} 
    Yang, S., Khare, K., Lin, P.C., 2010.
    Harnessing surface wrinkle patterns in soft matter.
    Adv. Funct. Mater. 20, 2550--2564.
    
    \bibitem[Yin et al., 2010]{Yin2010AB}
    Yin, J., Gerling, G.J., Chen, X., 2010.
    Mechanical modeling of a wrinkled fingertip immersed in water.
    Acta Biomater. 6, 1487--1496.
    
    \bibitem[Zhang et al., 2012]{Zhang2012JACS} 
    Zhang, Z., Zhang, T., Zhang, Y.W., Kim, K., Gao, H., 2012. 
    Strain-controlled switching of hierarchically wrinkled surfaces between superhydrophobicity and superhydrophilicity.
    J. Am. Chem. Soc. 28(5), 2753--2760.
    
    \bibitem[Zhao et al., 2020]{Zhao2020Langmuir}
Zhao, X., Wang, J., Huang, J., Li, L., Liu, E., Zhao, J., Li, Q., Zhang, X., Lu, C., 2020.
Path-guided hierarchical surface relief gratings on azo-films induced by polarized light illumination through surface-wrinkling phase mask.
Langmuir 36, 2837--2846.
    
    \bibitem[Zhou et al., 2022]{Zhou2022MMS}
    Zhou, M.T., Cai, Z.X., Fu, Y.B., 2022.
    Post-buckling of an elastic half-space coated by double layers.
    Math. Mech. Solids. 27(2), 193-209.
        
\end{thebibliography}

\end{document}